\documentclass[prb, twocolumn, showpacs,superscriptaddress]{revtex4-1}

\usepackage{graphicx}  
\usepackage{dcolumn}   
\usepackage{bm}        
\usepackage{amssymb}   

\newcommand{\qp}{\mathrm{qp}}
\newcommand{\tss}{\tau_\mathrm{ss}}
\newcommand*{\citen}[1]{%
	\begingroup
	\romannumeral-`\x 
	\setcitestyle{numbers}%
	\cite{#1}%
	\endgroup   
}

\begin{document}

\title{Measurement and Control of Quasiparticle Dynamics in a Superconducting Qubit}

\author{C. Wang}
\email{chen.wang@yale.edu}
\affiliation{Department of Applied Physics and Physics, Yale University, New Haven, CT 06520, USA}
\author{Y. Y. Gao}
\affiliation{Department of Applied Physics and Physics, Yale University, New Haven, CT 06520, USA}
\author{I. M. Pop}
\affiliation{Department of Applied Physics and Physics, Yale University, New Haven, CT 06520, USA}
\author{U. Vool}
\affiliation{Department of Applied Physics and Physics, Yale University, New Haven, CT 06520, USA}
\author{C. Axline}
\affiliation{Department of Applied Physics and Physics, Yale University, New Haven, CT 06520, USA}
\author{T. Brecht}
\affiliation{Department of Applied Physics and Physics, Yale University, New Haven, CT 06520, USA}
\author{R. W. Heeres}
\affiliation{Department of Applied Physics and Physics, Yale University, New Haven, CT 06520, USA}
\author{L. Frunzio}
\affiliation{Department of Applied Physics and Physics, Yale University, New Haven, CT 06520, USA}
\author{M. H. Devoret}
\affiliation{Department of Applied Physics and Physics, Yale University, New Haven, CT 06520, USA}
\author{G. Catelani}
\affiliation{Peter Gr\"unberg Institut (PGI-2), Forschungszentrum J\"ulich, 52425 J\"ulich, Germany}
\author{L. I. Glazman}
\affiliation{Department of Applied Physics and Physics, Yale University, New Haven, CT 06520, USA}
\author{R. J. Schoelkopf}
\affiliation{Department of Applied Physics and Physics, Yale University, New Haven, CT 06520, USA}

\date{\today}

\begin{abstract}
Superconducting circuits have attracted growing interest in recent years as a promising candidate for fault-tolerant quantum information processing.  Extensive efforts have always been taken to completely shield these circuits from external magnetic field to protect the integrity of superconductivity.  Surprisingly, here we show vortices can improve the performance of superconducting qubits by reducing the lifetimes of detrimental single-electron-like excitations known as quasiparticles.  Using a contactless injection technique with unprecedented dynamic range,  we quantitatively distinguish between recombination and trapping mechanisms in controlling the dynamics of residual quasiparticles, and show quantized changes in quasiparticle trapping rate due to individual vortices.  These results highlight the prominent role of quasiparticle trapping in future development of superconducting qubits, and provide a powerful characterization tool along the way. 
\end{abstract}


\maketitle

\section{Introduction}
Superconducting quantum circuits have made rapid progress~\cite{Devoret2013} in realizing increasingly sophisticated quantum states~\cite{Hofheinz2009, Vlastakis2013} and operations~\cite{Steffen2013, Chow2014} with high fidelity.  Excitations of the superconductor, or quasiparticles (QP), can limit their performance by causing relaxation and decoherence, with the rate approximately proportional to the QP density~\cite{Catelani2011}.  Operating at 20 mK or lower temperature, superconducting aluminum in thermal equilibrium should have no more than one pair of quasiparticles for the volume of the Earth.  However, a substantial background of quasiparticles has been observed in various devices from single electron~\cite{Knowles2012} or Cooper-pair transistors~\cite{Aumentado2004}, kinetic inductance detectors~\cite{Day2003, deVisser2014} to superconducting qubits~\cite{Sun2012, Riste2013, Pop2014}.  A detailed understanding of the generation mechanism and dominant relaxation processes will eventually be necessary to suppress this small background of quasiparticles and continue the improvement of these devices.  

Quasiparticle dynamics has been traditionally characterized by the ``lifetime" of excess quasiparticles relaxing towards a steady state, $\tss$.  A variety of techniques have been used to measure $\tss$ in aluminum, including low-frequency sub-gap electrical~\cite{Gray1971, Wilson2001} or thermal transport~\cite{Ullom2000, Rajauria2012}, resonance frequency shifts~\cite{Barends2008, deVisser2011} and more recently, qubit energy decay~\cite{Lenander2011, Wenner2013}.  Electron-phonon mediated pair-recombination has been established as the canonical mechanism of QP decay~\cite{Kaplan1976}.   Single-quasiparticle loss mechanisms in the presence of quasiparticle ``traps" such as normal metal contacts~\cite{Goldie1990, Joyez1994, Ullom2000, Rajauria2012, Knowles2012}, engineered gap inhomogeneity~\cite{Friedrich1997, Aumentado2004, Court2008}, Andreev bound states~\cite{Levenson-Falk2014} or magnetic field penetration~\cite{Ullom1998, Peltonen2011} have also been studied.

Ascertaining the relative importance of quasiparticles to the decoherence of any superconducting qubit, versus other mechanisms such as dielectric loss or radiation, remains challenging.  Ideally, one would like to probe and vary QP dynamics in a highly-coherent qubit, without requiring additional circuit elements or constraints that compromise its performance.  Here we introduce such a technique, capable of measuring the QP dynamics of a qubit in operation and quantifying the processes of recombination, trapping, diffusion and background generation of quasiparticles.  With this technique, we directly measure the trapping of quasiparticles by a single vortex, a basic property of superconductors.  We demonstrate that QP trapping by vortices can suppress the background QP density, resulting in surprising net improvement of qubit coherence in spite of the well-known vortex flow dissipation~\cite{Bardeen1965}.

In this Article, we present time-domain measurements of quasiparticle relaxation in 3D transmon qubits~\cite{Paik2011} over 2-3 orders of magnitude in density.  We find the QP dynamics can be dominated by either recombination or trapping effects depending on the device geometry and the resultant presence or absence of vortices.  
We demonstrate strong \textit{in-situ} control of QP dynamics by magnetic field, and measure an intrinsic single-vortex trapping ``power" of $(6.7\pm0.5)\times10^{-2}$ cm$^2$s$^{-1}$ (\textit{i.e.} $\tss$ = 1 s induced by a single vortex over an area of $6.7\times10^{-2}$ cm$^2$).   Improvements of relaxation time ($T_1$) and coherence time ($T_{2E}$) by more than a factor of 2 are observed in one geometric design of devices when the devices are cooled in a small magnetic field (10-200 mG).  We measure a stray QP generation rate of about $1\times10^{-4}$s$^{-1}$, suggesting the long coherence time of the widely-adapted design of 3D transmons~\cite{Paik2011, Sears2012, Riste2013, Vijay2012} may already be greatly assisted by unintentional vortices.  Improved coherence by field-cooling, also being reported in superconducting planar resonators~\cite{Nsanzineza2014} and fluxonium qubits~\cite{Vool2014} during the preparation of our manuscript, can be definitively correlated with QP trapping rates of vortices as probed by our measurements of QP dynamics.

\section{Results}

\begin{figure}[tbp]
    \centering
    \includegraphics[width=0.49\textwidth]{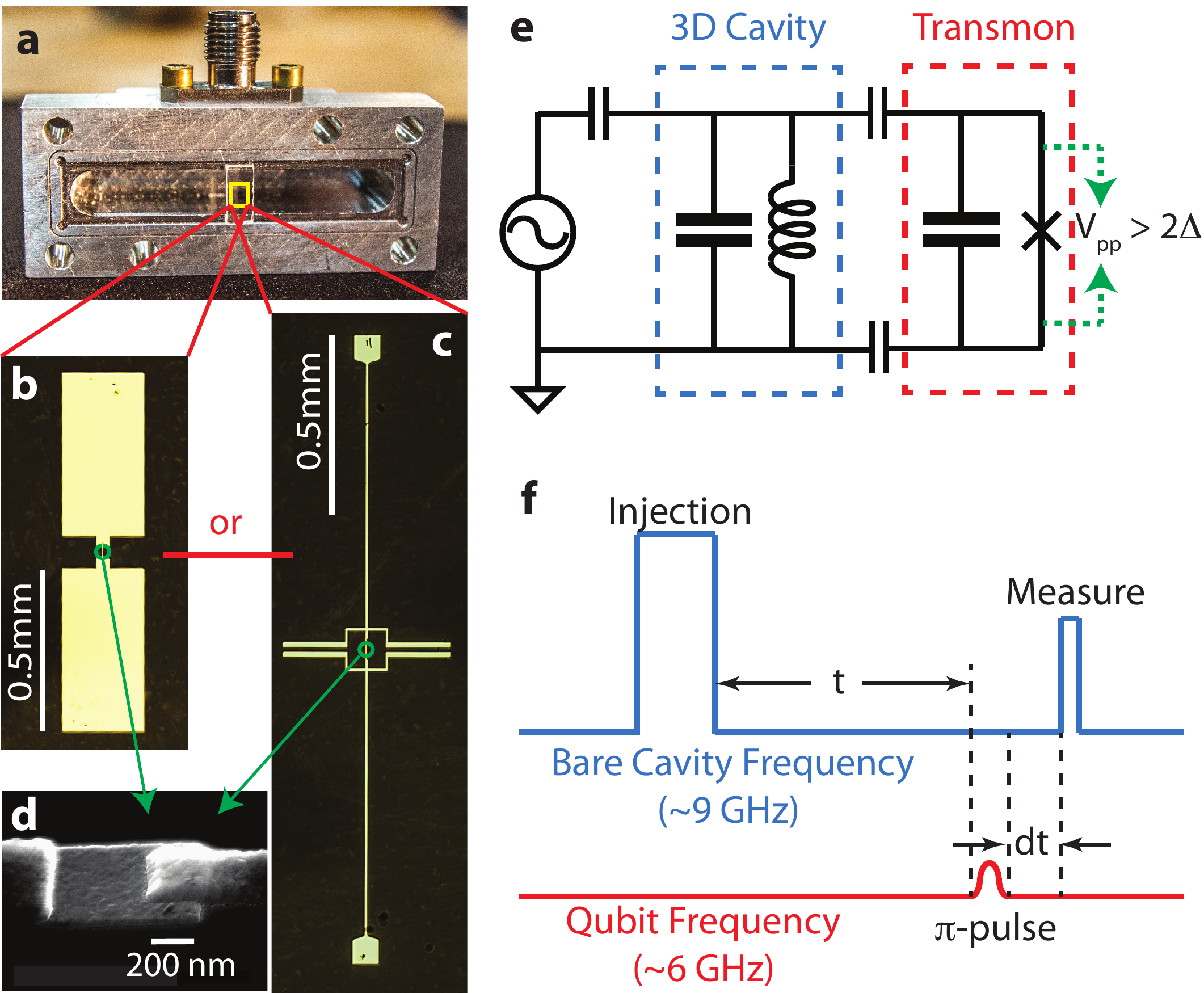}
    \caption{\textbf{Device and measurement schematics.} \textbf{a}, Photograph image of a 3D aluminum cavity loaded with a transmon qubit. \textbf{b}, \textbf{c}, Optical images of a Type A device and a Type B device. \textbf{d}, Scanning electron microscope image of a Josephson junction located in between the two electrodes of a transmon. \textbf{e}, Effective circuit of the cavity-qubit system.  Injection of quasiparticles is achieved by applying a resonant microwave voltage to the 3D cavity capacitively coupled to the junction. \textbf{f}, Pulse sequences for measuring the decay of quasiparticles based on qubit $T_1$.  A $\pi$-pulse is applied to excite the qubit from $|g\rangle$ to $|e\rangle$ at a variable delay $t$ after the injection pulse, followed by a readout pulse after another variable delay $dt$ $(dt\ll t)$.}
    \label{injection}
\end{figure}

\textbf{Injection and measurement of quasiparticles.}
In our experiment, we inject quasiparticles into two types of transmon qubits in a 3D circuit QED architecture~\cite{Paik2011} using only the existing microwave ports.  Each transmon qubit comprises a single Al/AlO$_x$/Al Josephson junction shunted by a large Al coplanar capacitor (electrodes) on a c-plane sapphire substrate.  Type A devices are very similar to those in ref.~\citen{Paik2011} with a pair of large 500 $\mu$m $\times$ 250 $\mu$m electrodes (Fig.~\ref{injection}b).  The electrodes of Type B devices are composed of a narrow (6 to 30 $\mu$m wide) coplanar gap capacitor and a pair of 80 $\mu$m $\times$ 80 $\mu$m ``pads" (see Fig.~\ref{injection}c and Supplementary Note 1), but provide total capacitance and qubit Hamiltonian parameters very similar to Type A devices.  Chips containing one or two qubits are mounted in 3D aluminum or copper rectangular waveguide cavities (Fig.~\ref{injection}a) and all measurements are done in an Oxford cryogen-free dilution refrigerator at base temperature of 15-20 mK, with magnetic field shielding, infrared shielding and filtering described in ref.~\citen{Sears2012}.  To inject quasiparticles, similarly to ref.~\citen{Vool2014} we apply a high-power microwave pulse at the bare cavity resonance frequency from the input port. The injection pulse creates about $10^5$ circulating photons in the cavity, resulting in RF voltage across the Josephson junction that exceeds the superconducting gap, and produces $\sim10^5$ quasiparticles per $\mu$s.  The duration of the injection pulse is long enough (200-500 $\mu$s) so that the injected quasiparticles can fully diffuse within the device while the production and loss of quasiparticles reach a dynamic balance. (See Supplementary Notes 2 and 3 for analysis of QP injection and diffusion.)

\begin{figure*}[tbp]
	\centering
    	\includegraphics[width=0.55\textwidth]{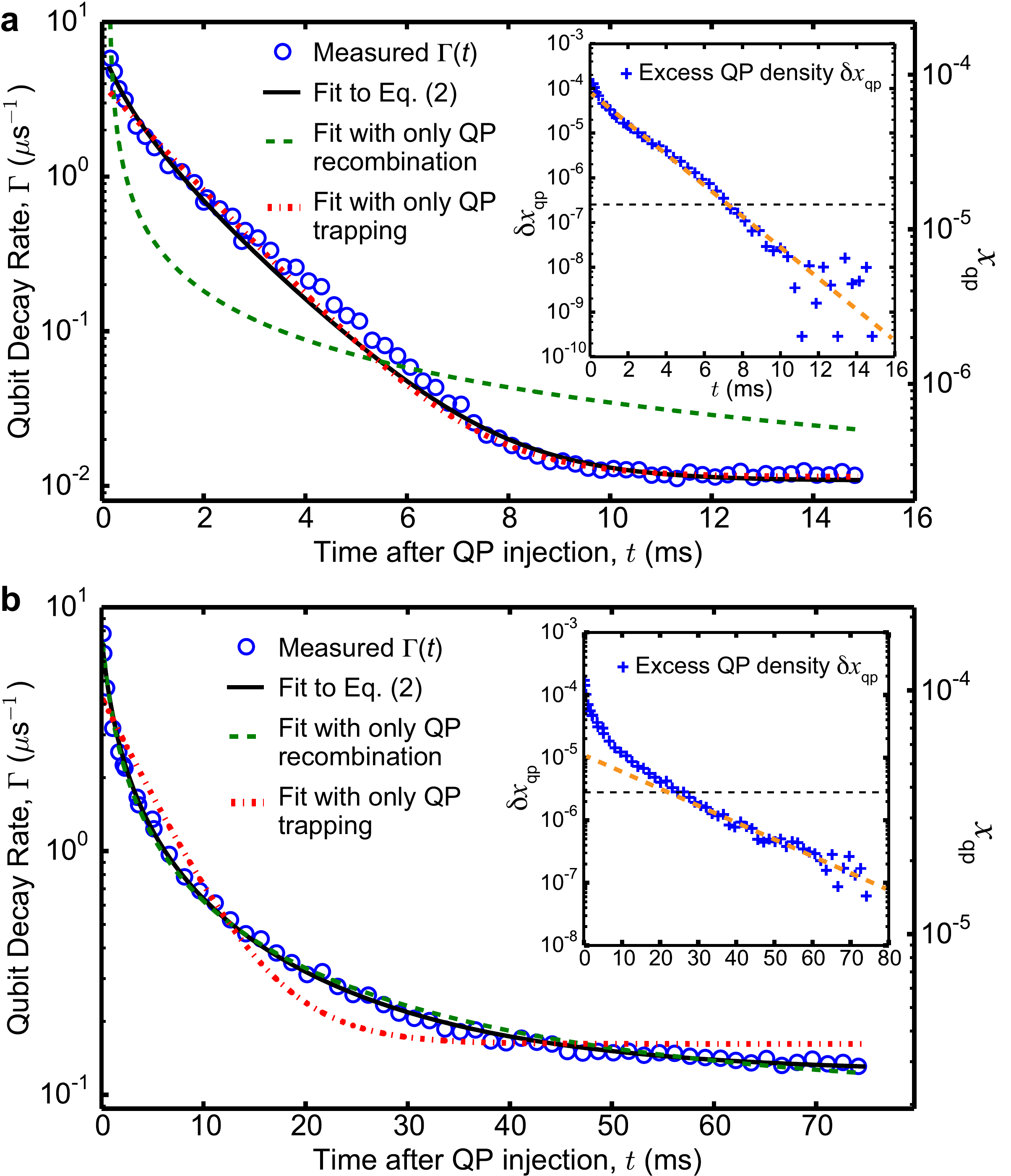}
    	\caption{\textbf{Quasiparticle decay dynamics characterized by the recovery functional forms of qubit decay rate.}  Qubit energy decay rate $\Gamma$ as a function of time $t$ after quasiparticle injection pulses and fits to various functional forms for Device A1 (\textbf{a}, dominated by trapping) and Device B1 (\textbf{b}, dominated by recombination) in separate aluminum 3D cavities.  For illustration purposes, we label $x_\qp=\Gamma/C$ on the right axes
, neglecting qubit relaxation from non-QP mechanisms that have a stringent bound $\Gamma_{ex}\leq\Gamma_0$.  Insets, replots of the $\Gamma(t)$ data in main panels in the form of excess QP density $\delta x_\qp$ after subtracting a background (whose magnitude $\Gamma_0/C$ is shown by the black dashed line and is an upper bound of $x_0$).  The magenta dashed line is guide to the eye of the exponential decay of $\delta x_\qp$ approaching the steady state.}
	\label{decayfit}
\end{figure*}

We use the recovery of the energy relaxation time ($T_1$) of the qubit as a direct and calibrated probe of the decay of QP density.  Standard microwave pulse sequences are applied to determine the qubit $T_1$ following a variable delay $t$ after the QP injection (Fig.~\ref{injection}f and Supplementary Note 4), from which we extract the qubit relaxation rate $\Gamma=1/T_1$ as a function of $t$ as shown in Fig.~\ref{decayfit}.  Despite possible heating due to the injection pulse, we find the effective temperature of the qubit and quasiparticle bath does not exceed 70 mK for the entire range of our measurement (Supplementary Note 5), therefore thermal generation of quasiparticles and spontaneous $|g\rangle\to|e\rangle$ transition of the qubit can be neglected.  We use $x_\qp$ to represent the QP density near the Josephson junction normalized by the Cooper pair density ($n_{cp}\approx4\times10^6$ $\mu$m$^{-3}$ for aluminium).  It is related to the measured qubit decay rate by $\Gamma(t)=Cx_\qp(t) + \Gamma_{ex}$, where $\Gamma_{ex}$ is a constant qubit decay rate due to non-quasiparticle dissipation mechanisms, and $C=\sqrt{2\omega_q\Delta/\pi^2\hbar}$ is a calculated~\cite{Catelani2011} and confirmed~\cite{Paik2011} constant involving only the superconducting gap $\Delta(\approx180$ $\mu e$V) and the angular frequency of the transmon $\omega_q(\approx2\pi\cdot6$ GHz).  Since $\Gamma_{ex}$ is strictly bounded by the qubit relaxation rate without QP injection, $\Gamma_{ex}\leq\Gamma_0=\Gamma(t\to\infty)$, for a significant range of the data, QP density can be approximated by $x_\qp(t)\approx\Gamma(t)/C$.  

\textbf{Distinguishing between quasiparticle recombination and trapping.}
The dynamics of the quasiparticle density $x_\qp$ near the junction in the presence of recombination and trapping can be modeled by the following equation (see Supplementary Note 6 for details):
\begin{equation}
	\frac{dx_\qp}{dt}=-rx^2_\qp-sx_\qp+g
	\label{eq:dynamics}
\end{equation}
The quadratic term describes the canonical QP recombination in pairs with a recombination constant $r$.  The linear term describes trapping effects that localize or remove single quasiparticles from tunneling across the Josephson junction and inducing qubit relaxation.  The effective trapping rate $s$ depends not only on the property and density of the trapping sites, but also their geometric distribution and associated diffusion time scale.  The constant term $g$ describes QP generation rate by pair-breaking stray radiation or other unidentified sources~\cite{Martinis2009}.  If trapping is dominant ($s\gg rx_\qp$ for most of the measured range of $x_\qp$), the decay of $x_\qp$ follows an exponential function.  This is a surprisingly good approximation for the $\Gamma(t)$ we measured in Device A1 (of Type A) (red fit in Fig.~\ref{decayfit}a).  On the other hand, if recombination dominates, the decay of $x_\qp$ follows a hyperbolic cotangent function, with initially a steep $1/t$ decay crossing over to an exponential tail.  We measure $\Gamma(t)$ in Device B1 (of Type B) strikingly close to this limit (green fit in Fig.~\ref{decayfit}b).   

To analyze both recombination and trapping more quantitatively, we solve Eq.~(\ref{eq:dynamics}) analytically, yielding a four parameter fit for $\Gamma(t)$ (black curves in Fig.~\ref{decayfit}):
\begin{equation}
	\Gamma(t)=C x_i \frac{1-r'}{e^{t/\tss}-r'}+\Gamma_0
	\label{eq:fitting}
\end{equation}
where $x_i$ is the initial injected QP density, $\Gamma_0=C x_0+\Gamma_{ex}$ is the qubit relaxation rate without QP injection, consisting of contributions from both `background QP density $x_0$ and other mechanisms. $r'$ is a dimensionless fit parameter ($0<r'<1$).  Note that as $t\to\infty$, Eq.~(\ref{eq:fitting}) approaches an exponential decay with time constant $\tss$. The recombination constant $r$ and the trapping rate $s$ can be determined from these fit parameters (Supplementary Note 6).  For B1, a fit to Eq.~(\ref{eq:fitting}) gives an exponential tail with $\tss=18\pm2$ ms, a recombination constant $r=1/(170\pm20$ ns) and a weak trapping rate $s = 1/(30 \pm\,^{24}_{12}$
ms).  For A1, we find $s\approx 1/\tss=1/(1.5\pm0.1$ ms) and $r=1/(105\pm30$ ns), with the trapping term dominating most of the measurement range ($x_\qp<10^{-4}$).

The qualitative difference between the functional forms of QP decay in the two devices can be better illustrated by plotting the excess quasiparticle density due to QP injection, $\delta x_\qp=x_\qp(t)-x_0$, as a function of time (Fig.~\ref{decayfit} insets).  
For A1, the instantaneous QP decay rate indicated by the slope of $\delta x_\qp(t)$ remains equal to its steady-state value (1/$\tss$) even when the QP density is orders of magnitude higher than its background density (\textit{i.e.}~when $\delta x_\qp(t)\gg \Gamma_0/C \geq x_0$).  For B1, the slope increases significantly when $\delta x_\qp(t)>x_0$.

\begin{figure*}[tbp]
    \centering
    \includegraphics[width=0.94\textwidth]{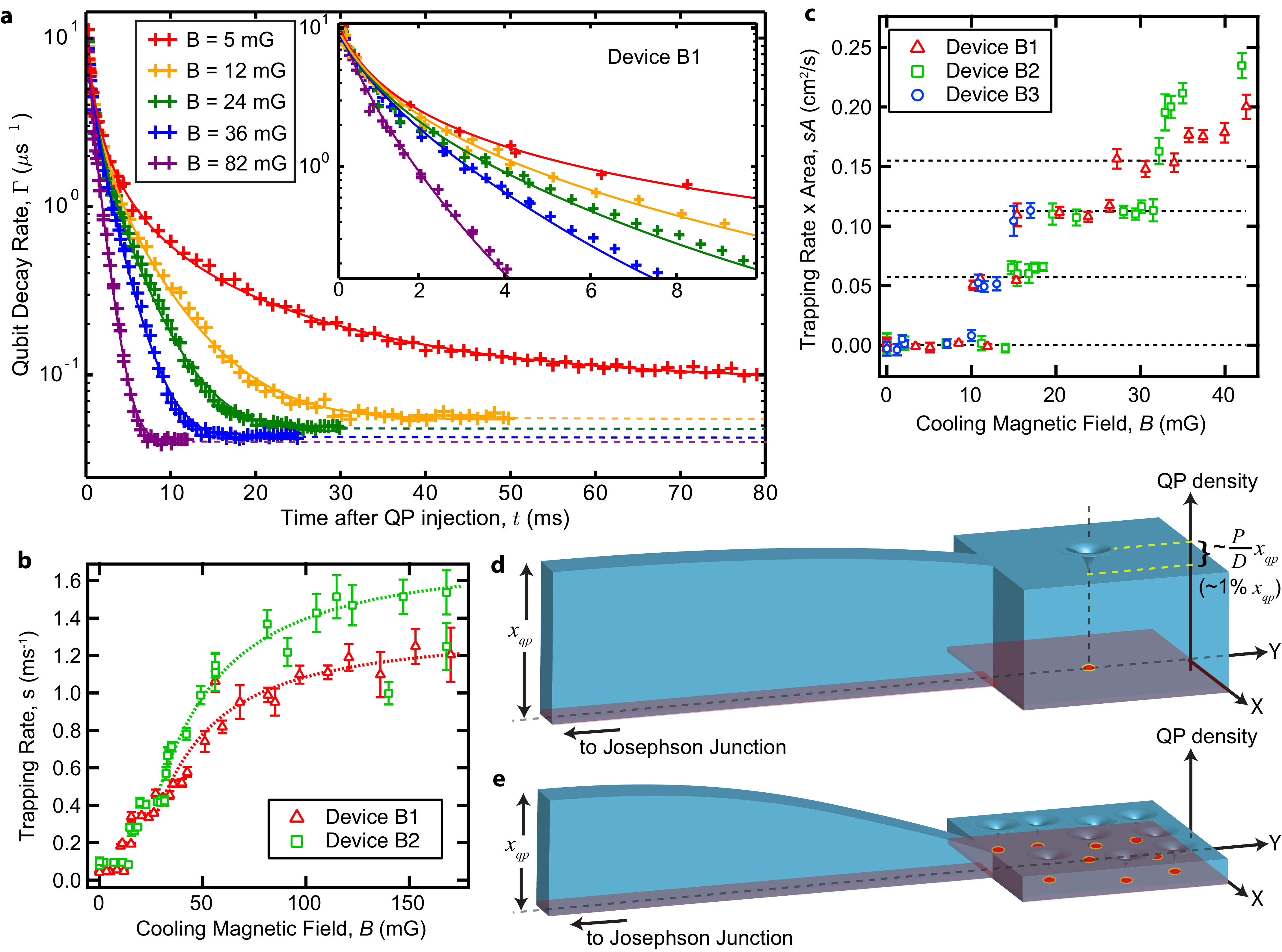}
    \caption{\textbf{Quasiparticle trapping by vortices.}  \textbf{a}, Qubit decay rate $\Gamma$ as a function of time $t$ after quasiparticle injection pulses for Device B1 cooled in a few selected magnetic fields $B$.  Solid lines fits to Eq.~(\ref{eq:fitting}).  Dashed lines, background qubit decay rate indicating the ``steady-state" $T_1$ of the qubit.  Inset, zoom-in of the initial part of the decay.  \textbf{b}, QP trapping rate $s$ as a function of cooling magnetic field $B$ for Devices B1 and B2.  Each point represents data acquired after one field-cool thermal cycle (from above the $T_c$ of aluminum), with the error bars showing the 1 s.d. of fluctuations for repetitive measurements within each thermal cycle.  Dotted lines, fits to the QP trapping/diffusion model.  \textbf{c}, QP trapping rate after subtracting a zero-vortex background and multiplying by the total area of the device, as a function of $B$ for Devices B1, B2 and B3.  Dashed lines are guides to the eye showing the discrete steps associated with 0, 1, 2 and 3 vortices.  \textbf{d}, Cartoon schematic (not to scale) of the near-homogeneous spatial distribution of QP density, represented by height of the blue volume (``tank water level"), from the narrow wire (left) to the trapping pad (right) in a Type B device when only one vortex is present.  QP density near the vortex displays a small ripple with a relative depth on the order of $P/D\approx 1\%$.  \textbf{e}, Cartoon schematic (not to scale) of the QP density distribution with large number of vortices in the pad.  The density at the trapping pad is much lower than $x_\qp$ near the junction.}
    \label{multifield}
\end{figure*}

\textbf{Controlling quasiparticle dynamics by cooling in magnetic field.}
Why do the two devices with identical material properties and similar qubit properties differ so much in quasiparticle relaxation dynamics? We attribute this to the trapping effect from vortices, regions with diminished superconducting gap, in the large electrodes in A1 despite the small residual field ($B\sim$ 1-2 mG) we are able to achieve with our magnetic shielding measures.  On the other hand, Device B1 is likely free of vortices due to the much narrower geometry of the electrodes.  To test this hypothesis, we repeat $\Gamma(t)$ measurements in B1 after cooling the device through the critical temperature ($T_c$) in a perpendicular magnetic field of variable magnitude $B$ in either polarity.  Indeed as $B$ increases, we observe significant acceleration of QP decay with increasingly pronounced single-exponential characteristics, indicating enhanced QP trapping (Fig.~\ref{multifield}a).  In comparison, changing the applied magnetic field at 20 mK does not produce measurable changes in quasiparticle dynamics.

By fitting $\Gamma(t)$ to Eq.~(\ref{eq:fitting}) at each cooling field, we find that: 1) the recombination constant $r$ remains unchanged within fitting uncertainty, 2) the trapping rate $s$ increases in discrete and near-equal steps for small magnetic fields ($B\lesssim40$ mG) (Fig.~\ref{multifield}c), 3) over a broader field range, $s$ increases approximately linearly with $B$ and saturates at $\gtrsim1$ ms$^{-1}$ at high field ($B\gtrsim100$ mG) (Fig.~\ref{multifield}b).

Our observed critical field threshold, $B_k$, where the trapping rate $s$ starts to increase, corresponds to the expected entry of the first vortex in one of the 80 $\mu$m $\times$ 80 $\mu$m pads.   $B_k$ can be estimated based on a thermodynamic analysis of a vortex in a thin superconducting disk~\cite{Kogan2004} together with consideration of vortex creation-annihilation kinetics~\cite{Kuit2008}, giving $B_k\sim8$ mG, close to our observed values of 11 mG for B1 in an Al cavity, 14 mG and 10 mG for other two devices of Type B, B2 and B3, in Cu cavities.  In comparison, $B_k$ for Device A1 is estimated to be about 0.5 mG, lower than the estimated residual field (and its possible inhomogeneity) of our setup and therefore cannot be observed experimentally.

\begin{figure*}[tbp]
    \centering
    \includegraphics[width=0.8\textwidth]{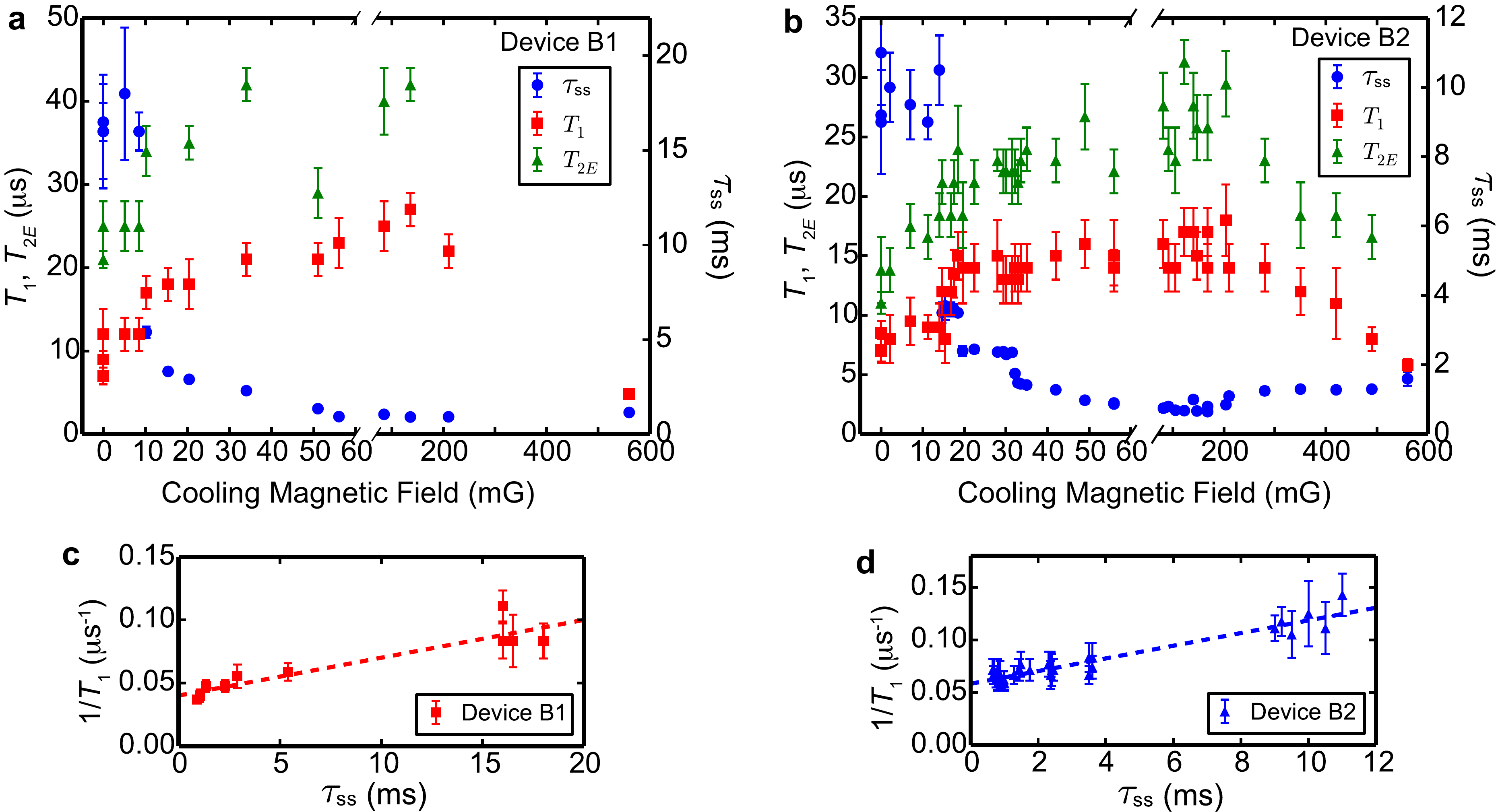}
    \caption{\textbf{Improved qubit coherence by cooling in magnetic field and its correlation with reduced QP lifetime.} \textbf{a}, \textbf{b}, Qubit steady-state relaxation time $T_1$, coherence time with echo $T_{2E}$ and quasiparticle lifetime $\tss$ as a function of cooling magnetic field $B$ for Devices B1 (\textbf{a}) and B2 (\textbf{b}).  Each point represents data acquired after one field-cool thermal cycle, with the error bars showing the 1 s.d. of fluctuations for repetitive measurements within each thermal cycle.  Note the discontinuous scale of the X-axes. \textbf{c}, \textbf{d}, $1/T_1$ versus $\tss$ as varied by cooling in different magnetic fields for B1 (\textbf{c}) and B2 (\textbf{d}).  The dashed lines are linear fits based on Eq.~(\ref{eq:t1tauqp}), giving $\Gamma_{ex}=1/(26$ $\mu$s), $g=0.7\times10^{-4}$s$^{-1}$ for B1 and $\Gamma_{ex}=1/(17$ $\mu$s), $g=1.3\times10^{-4}$s$^{-1}$ for B2 respectively.}
    \label{t1better}
\end{figure*}

\textbf{Quasiparticle trapping by individual vortices.}
The discrete trapping rates at small magnetic field are strongly suggestive of a fixed quasiparticle ``trapping power" for each individual vortex.  We define trapping power, $P$, by modeling a vortex as a point object at a 2D spatial coordinate $\vec{r}_0$ with a delta-function local trapping rate $P\delta(\vec{r}-\vec{r}_0)$.  $P$ could be more microscopically modeled as the product of QP trapping rate in the vortex core and an effective trapping area~\cite{Ullom1998, Nsanzineza2014}.  However, the ``trapping power" representation offers the advantage of a general formulation without invoking any microscopic models.  In the limit that diffusion is much faster than trapping, the total microscopic trapping power of $N$ vortices, $NP$, manifests itself macroscopically as the product of the measured trapping rate and the total area, $A$, of the device, \textit{i.e.} $sA=NP$.  For a small number of vortices,  we observe quantized changes of $sA$ products in steps of $\sim0.06$ cm$^2$s$^{-1}$ consistent between all three systematically-measured Type B devices with up to $50\%$ difference in device areas (Fig.~\ref{multifield}c).  In Fig.~\ref{multifield}c we have subtracted a relatively small (zero-vortex) background trapping rate that varies from device to device (zero-field $s<0.05$ ms$^{-1}$ for B1 but $\sim0.18$ ms$^{-1}$ for B3), whose origin remains to be explored in future studies.  Assuming each step corresponds to the entry of one vortex (which is stochastically most probable and also suggested by the widths of the steps), and adjusting for the finite speed of quasiparticle diffusion (Supplementary Note 6B), we determine trapping power $P=(6.7\pm0.5)\times10^{-2}$ cm$^2$s$^{-1}$ as an intrinsic property of each individual vortex in our superconducting film.

The reduced step heights and the eventual saturation of $s$ at higher magnetic field can be fully explained by QP diffusion using realistic geometric parameters of our device (Supplementary Note 6B).  When there are a large number of vortices in the pads, the apparent trapping rate $s$ is limited by the diffusion time for quasiparticles to reach the trapping pad from other regions of the device (Fig.~\ref{multifield}e).  The saturated trapping rate is higher for B2 because of the smaller volume of its gap capacitor.  By fitting $s$ as a function of $B$ over a large range (Fig.~\ref{multifield}b) for both devices, we determine the diffusion constant $D=18\pm2$ cm$^2$s$^{-1}$ for our Al film at 20 mK, consistent with the values measured in X-ray single-photon spectrometers adjusted for different temperature~\cite{Friedrich1997}. 

\textbf{Quasiparticle recombination constant.}
Across seven devices (Supplementary Table 1) we measure recombination constants $r$ in the range of 1/(170 ns) to 1/(80 ns), consistent with the theoretical electron-phonon coupling time of aluminum $\tau_0=$ 438 ns~\cite{Kaplan1976} adjusting for the phonon trapping effect~\cite{Rothwarf1967}.  Recombination-generated phonons can re-break Cooper-pairs before escaping into the substrate, reducing the effective recombination constant by a factor $F$, giving $r=4(\frac{\Delta}{k_BT_c})^3/(F\tau_0)=21.8/(F\tau_0)$.  For our 80 nm bi-layer Al film on sapphire the best estimate of $F$ is in the range of 5 to 10 considering the strong acoustic mismatch at the interface~\cite{Kaplan1979}.  The reported values of $r$ for Al in the literature range from 1/(120 ns) to 1/(8 ns)~\cite{Gray1971, Wilson2004, Ullom1998, Barends2008, deVisser2011}.  The recombination constant we measured directly from the power-law decay characteristics is near the low end.  It is comparable to $r=1/(120$ ns) extracted from DC steady-state injection measurements in extended Al films with similar thickness on sapphire\cite{Gray1971}. 

Our measured value of $r$ suggests the effect of recombination is extremely weak for the sparse QP background in typical superconducting qubits (recombination-resulted $\tss\sim50$ ms for $x_\qp\sim10^{-6}$).  Even one single vortex (occupying less than one-millionth of the total area of a Type B device) can eliminate quasiparticles much faster than the intrinsic recombination process, as demonstrated by the drastically different QP decay curves for 0 (red), 1 (orange) and 2 (green) vortices in Fig.~\ref{multifield}a.

\textbf{Improved qubit coherence by vortices.}
In strong correlation with the reduced QP lifetime due to vortex trapping, we observe dramatic improvement of qubit coherence as a result of the suppressed background QP density.  Improved qubit $T_1$ is already demonstrated by the lower background $\Gamma_0$ in Fig.~\ref{multifield}a at higher cooling field.  This ``steady-state" $T_1$ of the qubit can also be measured separately without QP injection, which more than doubled from its zero-field value over a wide range of cooling magnetic field for both Devices B1 (Fig.~\ref{t1better}a) and B2 (Fig.~\ref{t1better}b).  The coherence time with Hahn echo, $T_{2E}$, shows similar relative improvement because it is close to the limit of $2T_1$.  

We can separate qubit loss mechanisms based on the linear relation between $1/T_1$ and $\tss$ (noting $x_0\approx g\tss$), 
\begin{equation}
	\frac{1}{T_1}=C g \tss + \Gamma_{ex}
	\label{eq:t1tauqp}
\end{equation}
where the intercept equals the qubit relaxation rate due to other loss mechanisms, and the slope reveals the stray QP generation rate $g$ (Fig.~\ref{t1better}c, d).  In both B1 and B2, $\Gamma_{ex}$ is likely dominated by dielectric loss at the Al/Al$_2$O$_3$ interface under the gap capacitor~\cite{Wenner2011}.  We find QP generation rate $g\approx0.7$-$1.3\times10^{-4}$s$^{-1}$ in the two devices over several thermal cycles, or 0.3-0.6 quasiparticles created per ms for every $\mu$m$^3$ of volume.   

\textbf{Vortex flow dissipation.}
The eventual decrease of $T_1$ and $T_{2E}$ at high magnetic field ($B\gtrsim200$ mG) can be attributed to vortices entering the gap capacitors where current density is high.  Such dissipation due to vortex flow resistance has been well-known to degrade the quality factor of superconducting resonators and qubits~\cite{Wang2009, Song2009}.  Therefore precautions such as multilayer magnetic shielding, specialized non-magnetic hardware and honeycomb-style device designs have been widely employed in the community to avoid vortices.  However, here we have shown that the benefit of vortices in suppressing non-equilibrium quasiparticles is not only relevant to practical devices, but can also significantly outweigh its negative impact if the locations of the vortices are optimized.

In device A1 (or most Type A devices, see Supplementary Table 1), our measured $\tss$ at nominally zero field implies the presence of about 20 vortices based on our measured single-vortex trapping power, consistent with its geometry assuming a residue magnetic field of 1-2 mG.  Introducing more vortices to Type A devices by cooling in an applied magnetic field can further reduce $\tss$, but no clear improvement of qubit $T_1$ is observed.  We attribute it to the limited range that $\tss$ can be varied (from its already small value at nominally zero field), and the vortex flow dissipation which clearly reduces $T_1$ at $B\gtrsim30$ mG (Supplementary Note 8).  Assuming both types of devices have similar $g$ as they are shielded and measured in the same setup, it is plausible to infer that the long coherence times of the widely-adopted Type A 3D transmons would have been limited to much lower values without the assistance by QP trapping of unintentional vortices.

\section{Discussion}

In our work, for the first time, the interaction between quasiparticles and a single vortex is measured.  The single-vortex trapping power ($P$) is an intrinsic property of an aluminum film, and has the same dimension as the diffusion constant ($D$).  The fact that we measure $P/D\approx 10^{-2}$ implies that a quasiparticle can diffuse through a vortex with only a small ($\sim1\%$) probability to be trapped.  Despite a vortex being a topological defect with $\Delta=0$ at its core, the spatial distribution of QP density is barely perturbed by the presence of a vortex, just like a small ``ripple" on the order of $1\%$ deep in a flat ``sea" (Fig.~\ref{multifield}d, see Supplementary Note 6E for calculation).   For a uniform film extended in 2D space, this relatively homogeneous QP distribution holds for all practical length scales at any density of vortices, so the trapping rate $s$ can be simply computed from the total trapping power.  In the high magnetic field limit, this leads to $s\propto B$ with a linear coefficient $P/\phi_0=0.3$ $\mu$s$^{-1}$G$^{-1}$ based on our measured $P$, close to the fitted value of 0.5 $\mu$s$^{-1}$G$^{-1}$ in ref.~\citen{Ullom1998}.   

Understanding and controlling the steady-state QP lifetime ($\tss$) may be important for a variety of superconducting devices. For superconducting qubits and resonators, achieving short $\tss$ (either with vortices or potentially more efficient methods such as band-gap engineering or normal metal traps) to suppress the background QP density is desirable. In other devices such as kinetic inductance detectors~\cite{Day2003, deVisser2014}, it may be desirable to obtain a long QP lifetime.  Our measured $\tss$ of 18 ms in a Type B device is so far the longest reported in aluminum.  Previous experiments in aluminum films showed that the expected exponential increase of $\tss$ with lower temperature saturates below 200 mK~\cite{Gray1971, Wilson2001, Barends2008} to about 3 ms at most~\cite{deVisser2014}.  

To shed light on the mechanisms limiting $\tss$ in this regime of extremely low QP density (without intentional QP traps), our technique allows quantitative separation between QP recombination and any residual trapping effects.  Our quantification of the weak recombination of background quasiparticles and the single-vortex trapping extends our understanding of QP dynamics into the 10's of millisecond regime, and thus sets a more stringent bound on possible additional mechanisms in limiting the QP lifetimes.

To distinguish between recombination and trapping of quasiparticles, our analysis relies critically on large dynamic range in $x_\qp$ to achieve sufficient contrast between the functional forms of QP decay.  A measurement near a steady state would observe the exponential tail of the decay, giving $\tss=1/(s+2rx_0)$ without distinguishing between the two mechanisms.  Such analysis of linear response has been traditionally carried out in time-domain measurements after photon pulses\cite{Goldie1990, Friedrich1997, Wilson2001, deVisser2011}.  Noise spectroscopy measurements~\cite{Wilson2001, deVisser2011} by design are also limited to measuring the single time scale of $\tss$.  The experiment of Lenander et al.~\cite{Lenander2011} introduced the use of qubit $T_1$ to probe quasiparticle dynamics, but the achieved dynamic range was below a factor of 4.  The dynamic range of 2-3 orders of magnitude in our experiment has been made possible by a long $T_1$ time in 3D transmons, the effectiveness of our microwave injection technique, and the geometric simplicity of an isolated aluminum island to eliminate out-diffusion.

A significant stray QP generation rate of about $1\times10^{-4}$s$^{-1}$ has been measured in our study.  Due to the large device volume and relatively long integration time, our measured $g$ should be considered a spatial-temporal average of QP generation rate.  Quite remarkably, it agrees within a factor of 3 with the average QP generation rate in a much smaller fluxonium qubit~\cite{Vool2014} with a much shorter $\tss$ where there is evidence for the discreteness of QP numbers and QP generation events.  This magnitude of QP generation rate, together with the weakness of recombination at low QP density, strongly suggests QP trapping should be an important ingredient for suppressing non-equilibrium $x_\qp$ in superconducting qubits and other devices to further improve performance.  Our method of extracting $g$ utilizing the sensitivity of Type B devices to QP generation should facilitate future identification of the QP generation source.  The injection and measurement technique introduced in this work can be readily applied to nearly all cQED implementations without any modification to device structure or measurement circuit, and can play a crucial role in the future development of quantum circuits as a powerful probe of quasiparticle dynamics. 

\section{Acknowledgements}
We acknowledge helpful discussions with D.~E.~Prober and B.~L.~T.~Plourde.  Facilities use was supported by YINQE and NSF MRSEC DMR 1119826.  This research was supported by IARPA under Grant No.~W911NF-09-1-0369, ARO under Grant No.~W911NF-09-1-0514, DOE Contract No.~DEFG02-08ER46482 (L.G.), and the EU under REA grant agreement CIG-618258 (G.C.).  Y.Y.G. acknowledges support from an A*STAR NSS Fellowship.

\section{Author Contributions} 
C.W. and Y.Y.G. carried out the experiment and performed data analysis based on the model developed by C.W., G.C.~and L.I.G.  The experimental method was conceived and developed by I.M.P., U.V., C.W., Y.Y.G., M.H.D.~and R.J.S.  ~C.A., T.B.~and R.W.H. provided further experimental contributions.  Devices are fabricated by C.W. and L.F.  ~C.W. and R.J.S. led the writing of the manuscript. All authors provided suggestions for the experiment, discussed the results and contributed to the manuscript.

\section{Additional Information}

\textbf{Competing financial interests:} The authors declare no competing financial interests.

\end{document}



\title{Supplementary Information for
``Measurement and Control of Quasiparticle Dynamics in a Superconducting Qubit"}



\author{C. Wang}
\email[]{chen.wang@yale.edu}
\affiliation{Department of Applied Physics and Physics, Yale University, New Haven, CT 06520, USA}
\author{Y. Y. Gao}
\affiliation{Department of Applied Physics and Physics, Yale University, New Haven, CT 06520, USA}
\author{I. M. Pop}
\affiliation{Department of Applied Physics and Physics, Yale University, New Haven, CT 06520, USA}
\author{U. Vool}
\affiliation{Department of Applied Physics and Physics, Yale University, New Haven, CT 06520, USA}
\author{C. Axline}
\affiliation{Department of Applied Physics and Physics, Yale University, New Haven, CT 06520, USA}
\author{T. Brecht}
\affiliation{Department of Applied Physics and Physics, Yale University, New Haven, CT 06520, USA}
\author{R. W. Heeres}
\affiliation{Department of Applied Physics and Physics, Yale University, New Haven, CT 06520, USA}
\author{L. Frunzio}
\affiliation{Department of Applied Physics and Physics, Yale University, New Haven, CT 06520, USA}
\author{M. H. Devoret}
\affiliation{Department of Applied Physics and Physics, Yale University, New Haven, CT 06520, USA}
\author{G. Catelani}
\affiliation{Peter Gr\"unberg Institut (PGI-2), Forschungszentrum J\"ulich, 52425 J\"ulich, Germany}
\author{L. I. Glazman}
\affiliation{Department of Applied Physics and Physics, Yale University, New Haven, CT 06520, USA}
\author{R. J. Schoelkopf}
\affiliation{Department of Applied Physics and Physics, Yale University, New Haven, CT 06520, USA}


\date{\today}

\pacs{}

\maketitle


\section{Details of Device Geometry}

\begin{figure}[bh]
    \centering
    \includegraphics[width=0.48\textwidth]{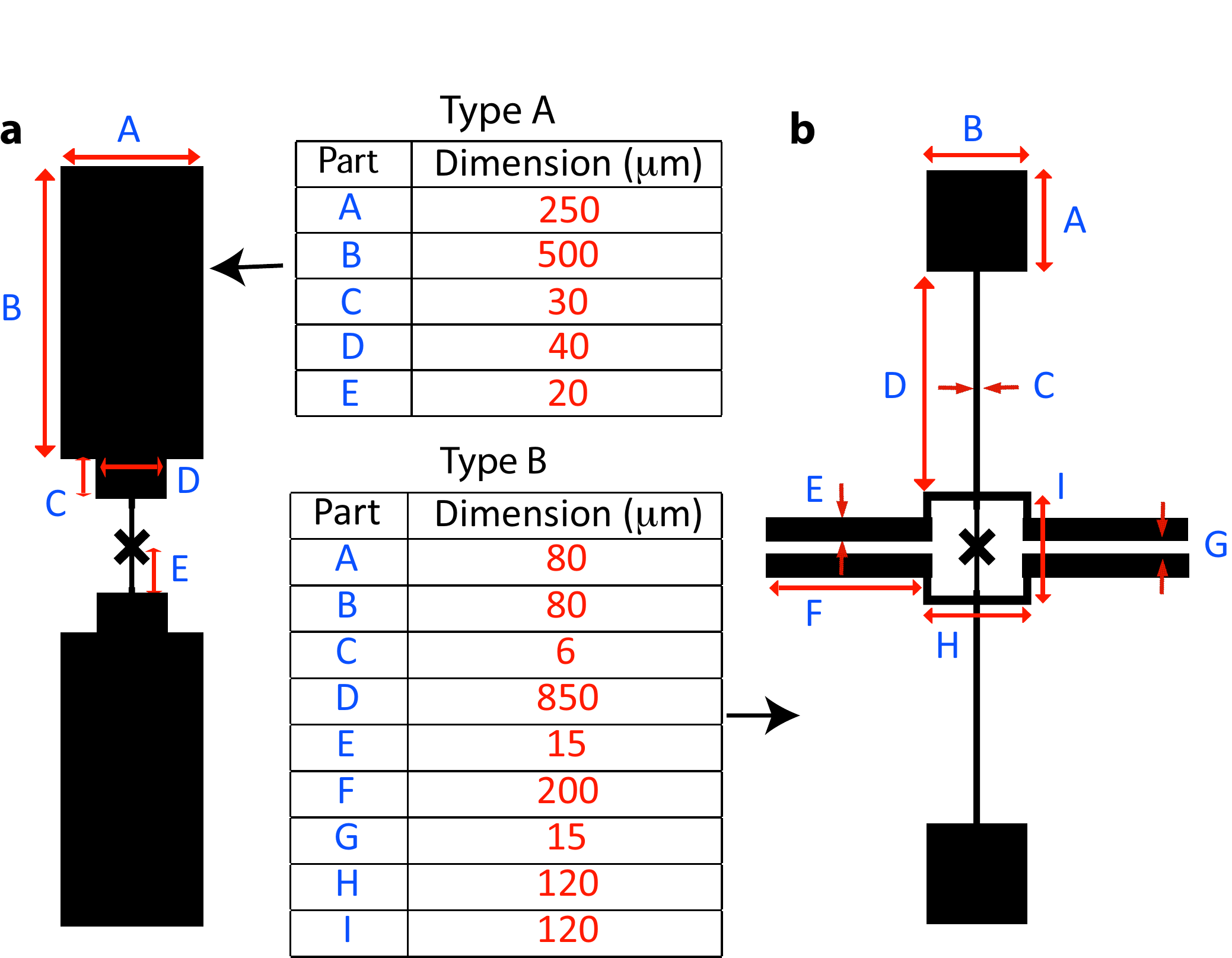}
    \caption{\textbf{Detailed device geometry.}  Dimensions of the electrodes of Type A (\textbf{a}) and Type B (\textbf{b}) transmons.}
    \label{geometry}
\end{figure}

The detailed dimensions of the electrodes of the two types of transmons are illustrated in Fig.~\ref{geometry}.  The width $E$ and distance $G$ of the coplanar gap capacitor in Type B are kept equal but varied from device to device.  All the other dimensions are fixed.  Device B1 has $E=G=15$ $\mu\mathrm{m}$, Device B2 has $E=G=10$ $\mu\mathrm{m}$, and Device B3 has $E=G=30$ $\mu\mathrm{m}$.

\section{Model of quasiparticle injection}

In our experiment, contactless injection of quasiparticles is realized by applying a high power microwave pulse at the bare cavity frequency to achieve an AC voltage $V_j>2\Delta/e$ across the Josephson junction.  The actual power used for injection is empirical without direct knowledge of $V$.  We find three regimes separated by two critical power thresholds: 1) low power: there is negligible microwave transmission at the bare cavity frequency, because with the presence of Josephson inductance the cavity mode frequency is shifted. 2) medium power: there is a drastic increase in transmission at the bare cavity frequency, and a small amount of QP's may be produced (affecting $\Gamma$ by up to a few times) with strong sample-to-sample variations.  3) high power: transmitted power becomes linear as a function of input power, and a large number of QP's are produced (affecting $\Gamma$ by up to 3 orders of magnitude).  The first critical transition in transmission power is the mechanism of the widely-used Jaynes-Cummings readout,\cite{Reed2010} and we believe the second transition in QP generation corresponds to $V_j\approx2\Delta/e$.  We typically use an injection power 1-4 dB above this second threshold power, which is about -70 to -55 dBm at the input port of the cavity based on our coarse estimate of the cable loss.  Here we provide a brief calculation of the microwave power required to realize $V_j=2\Delta/e$ based on a self-consistency argument.

Assuming the RMS voltage $V_j=2\Delta/e$,  the junction can be approximated by a normal resistor. We use a HFSS (Ansys) numerical package to simulate the quality factor of a lossless cavity containing our qubit structure with a tunnel junction resistance of $R_j$.  This ``internal" Q of the cavity found in the simulation is solely due to the dissipation at the normal-state junction resistor, which we denote by $Q_j$.  For our experimental parameter of $R_j\approx8$ k$\Omega$, we find $Q_j\approx1.1\times10^4$.

The power dissipated in the junction $P_j$ can be calculated as $P_j \approx V_j^2/R_j = 4\Delta^2/e^2R_j$.  $P_j$ is connected to the incoming power $P_{in}$ by:
\begin{equation}
	\frac{P_{j}}{P_{in}}=\frac{4Q^2_{tot}}{Q_{in}Q_j}
\end{equation}
where $Q_{tot}$ is the total Q of the cavity, composed of coupling Q's of the input and output microwave ports $Q_{in}$ and $Q_{out}$, the normal-state junction $Q_j$ and Q from the cavity wall (and dielectric, etc.)~$Q_w$:
\begin{equation}
	\frac{1}{Q_{tot}}=\frac{1}{Q_{in}}+\frac{1}{Q_{out}}+\frac{1}{Q_j}+\frac{1}{Q_w}
\end{equation}

\begin{figure*}[!htb]
    \centering
    \includegraphics[width=\textwidth]{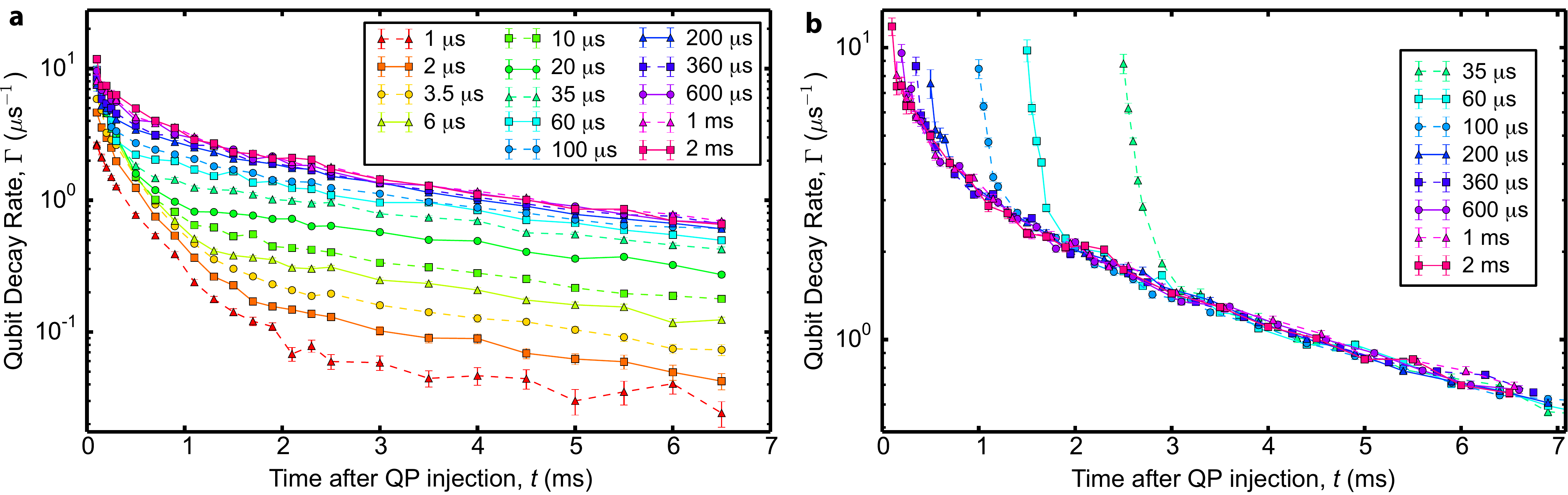}
    \caption{\textbf{Pulse-length-dependence of measured quasiparticle decay dynamics.} \textbf{a}, $\Gamma(t)$ for Device B2 cooled in zero magnetic field following quasiparticle injection pulses with a series of different pulse length $t_{inj}$.  A background qubit decay rate ($\Gamma_0$) has been subtracted.  \textbf{b}, Selected $\Gamma(t)$ traces from \textbf{a} with each trace shifted horizontally by a time offset.  Error bars represent one s.d. uncertainty in determining $T_1$ from individual exponential fits.}
    \label{pulselength}
\end{figure*}

For our devices in aluminum cavity, we typically have $Q_{in}\approx 2\times10^6$, $Q_{out}\approx 1\times10^5$, $Q_w\gg10^5$, so $Q_{tot}\approx Q_{j}$, therefore the input power required for $V_j=2\Delta/e$ is:
\begin{equation}
	P_{in}\approx \frac{Q_{in}}{4Q_j}P_{j}  \approx 200\frac{\Delta^2}{e^2R} \approx 10^{-9} \mathrm{W} =-60\mathrm{dBm}
\end{equation}

For our devices in copper cavity, we typically have $Q_{in}\approx 3\times10^5$, $Q_{out}\approx1.5\times10^4$, $Q_w\approx 1.5\times10^4$, so $Q_{tot}\approx4.5\times10^3$, so:
\begin{equation}
	P_{in}=\frac{Q_{in}Q_j}{4Q^2_{tot}}P_{j} \approx 200\frac{\Delta^2}{e^2R} \approx 10^{-9} \mathrm{W} =-60\mathrm{dBm}
\end{equation}
In both cases, we find the required input power similar to the injection power threshold we have observed.

So how many quasiparticles does the pulse produce?  We make another order-of-magnitude estimate here by assuming each tunneling electron produces one pair of QP's.  The number of tunneling electrons per second is given by $V_j/(R_je)$.  Therefore the QP injection rate is:
\begin{equation}
	G \approx 2\frac{V_j}{R_j e} \approx \frac{4\Delta}{e^3R_j} \approx 5\times10^5/\mu s
\end{equation}
A better estimate may consider that the instantaneous voltage across the junction is only above $2\Delta/e$ for part of an oscillation period, resulting in a slightly lower injection rate.

\section{Injection pulse length and the role of diffusion}

Since the injected quasiparticles are concentrated near the Josephson junction, their spatial distribution is in general not homogenous.  However, our measurement technique is only sensitive to the QP density near the junction.   (Note that we have explicitly specified the QP density for the near-junction area in the definition of $x_\qp$)  Therefore, quasiparticles diffusing away from the junction will also appear as a decay of $x_\qp$.  This complexity can be analyzed in depth by measurements of QP decay following injection pulses with different length.

In Fig.~\ref{pulselength}a we show a large series of $\Gamma(t)$ curves for Device B2 cooled in zero field following injection pulses with different pulse length $t_{inj}$.  A background qubit decay rate ($\Gamma_0$) has been subtracted so that the $\Gamma(t)$ shown here is strictly due to the excess quasiparticles from the injection.  We find:  1) For $t_{inj}\geq6$ $\mu\mathrm{s}$, all curves start from about the same $\Gamma$ at $t=100$ $\mu$s (the first point of our measurement), therefore extending injection pulse to more than $6\mu$s no longer increases QP density near the junction.  2) For 6 $\mu\mathrm{s}\leq t_{inj}\leq 600$ $\mu\mathrm{s}$, the $\Gamma(t)$ curves quickly deviate from each other after the initial points.  After a sufficiently long wait time $t$ (\textit{e.g.} see the last point of the curves $t=6.5$ ms), we see substantially higher $x_\qp$ for longer injection pulses.  This indicates a longer injection pulse injects more quasiparticles filling up remote areas of the device, while a shorter injection pulse results in a steeper initial decay due to quasiparticles diffusing away from the near-junction area.  3) As $t_{inj}\geq600$ $\mu\mathrm{s}$,  the $\Gamma(t)$ curves are completely indistinguishable.  In this regime, the QP injection has reached a dynamic balance with the QP diffusion and relaxation, which we call ``saturation injection", and the entire spatial distribution of QP density no longer changes with increasing $t_{inj}$.

By shifting each $\Gamma(t)$ curve in Fig.~S2a horizontally by a time offset, we find that all curves merge into a universal decay curve after some short-term behavior specific to each curve (Fig.~S2b) (which, as described above, is due to different initial spatial distribution of quasiparticles).  This universal curve is therefore the slowest eigenmode of the collective decay of quasiparticles in the entire electrodes, which is described by Eq.~(1) in the main text.   When recombination is dominant this mode is homogenous, and in the presence of trapping this mode may involve a steady spatial gradient of quasiparticles.   The time it takes for each measured $\Gamma(t)$ to converge to this decay mode depends on the diffusion time within the electrodes, but the deviation can be minimized by choosing an initial spatial distribution (after the injection pulse) as close to the eigenmode as possible.  Therefore, in order to use the intuitive Eq.~(1) in the main text without detailed simulation of the diffusion process, we choose $t_{inj}=360$ $\mu$s for the data presented in the main text for Device B2, close to the state of saturation injection. Furthermore, when analyzing $\Gamma(t)$ data using Eq.~(2) in the main text, we do not include data of $t<200$ $\mu$s, allowing sufficient additional diffusion time for the QP distribution to approach the slowest decay mode we are interested in.

From Fig.~S2 we can also estimate the time scale of diffusion in the device.  We observe saturation injection with pulse length $\gtrsim600$ $\mu$s.  In addition, we find that after a short injection pulse such as $t_{inj}=60$ $\mu$s it takes a similar amount of time $\sim500$ $\mu$s for $\Gamma(t)$ to merge into the universal decay curve (Fig.~S2b).  Furthermore, the two numbers are in very good agreement with the saturated trapping rate $1/(600$ $\mu$s) observed for this device [see Fig.~3b of the main text] which provides a reliable measure of the diffusion time scale (as modeled in Supplementary Note 6B).

\section{Measurement and data processing methods}

\begin{figure}[!htb]
    \centering
    \includegraphics[width=0.48\textwidth]{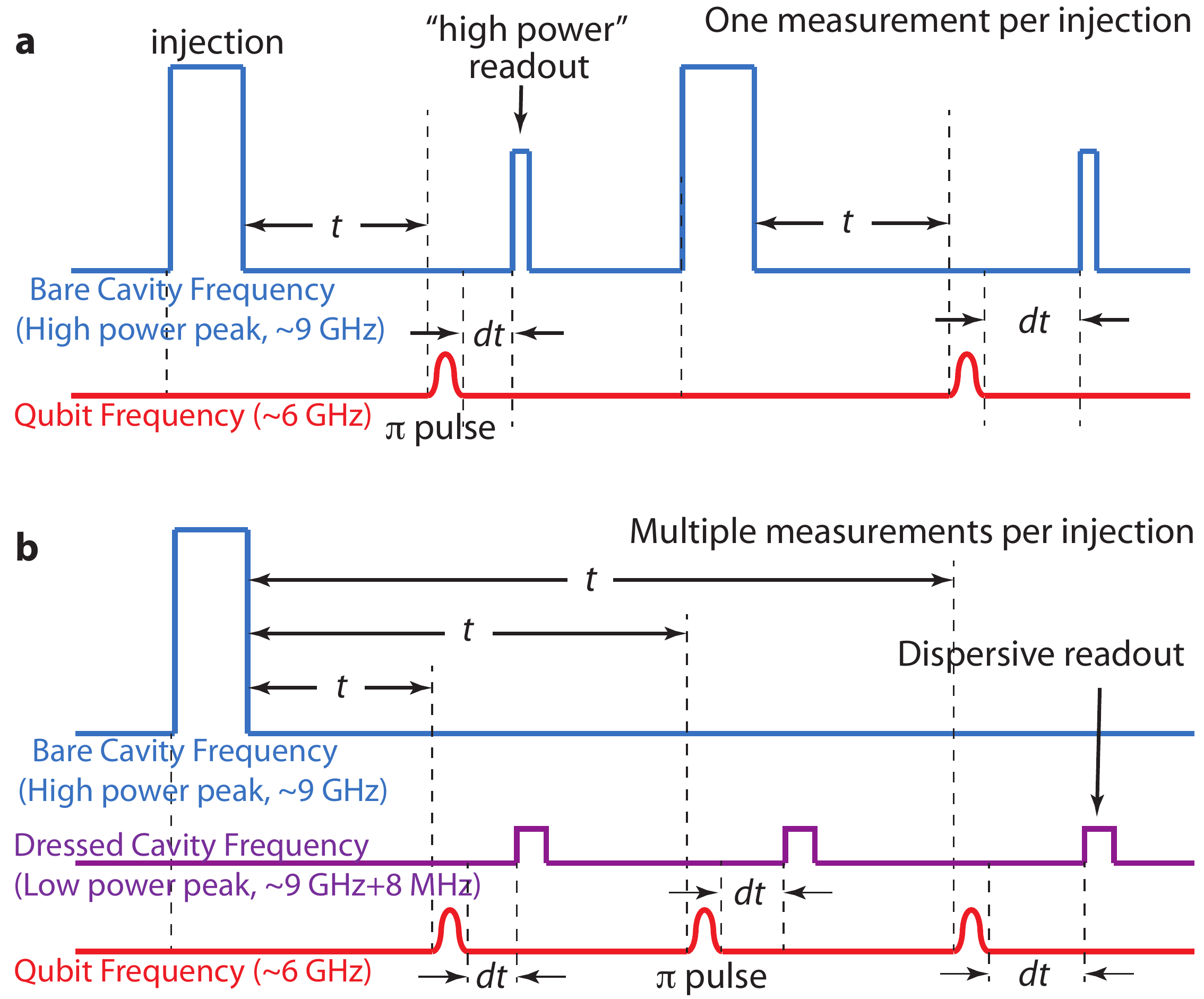}
    \caption{\textbf{Pulse sequences for qubit $T_1$ measurements.}  \textbf{a}, One measurement per injection using high power readout. \textbf{b}, Multiple measurement per injection using dispersive readout.}
    \label{pulseseq}
\end{figure}

We use a combination of dispersive readout and high-power readout (using the Jaynes-Cummings nonlinearity~\cite{Reed2010}) to measure $\Gamma(t)$ shown in the main text.  High-power readout provides much better signal-to-noise ratio (SNR), but we observe evidences that it may produce quasiparticles with mechanisms unclear (as we noted in Supplementary Note 2).  Therefore we only perform one readout following each injection pulse using the pulse sequence shown in Fig.~\ref{pulseseq}a.  Any unintended impact of the readout on the QP system should be completely erased by the next saturation injection (see Supplementary Note 3).  Dispersive readout has relatively poor SNR and requires much longer integration time, but it does not perturb quasiparticle dynamics significantly.  Therefore, one can perform multiple $T_1$ measurements after each injection pulse as shown in Fig.~\ref{pulseseq}b.  This is an efficient method when a very long delay time $t$ (10's of ms) is required for the measurement (\textit{e.g.}~when $\tau_\qp$ is long).

\begin{figure}[!htb]
    \centering
    \includegraphics[width=0.48\textwidth]{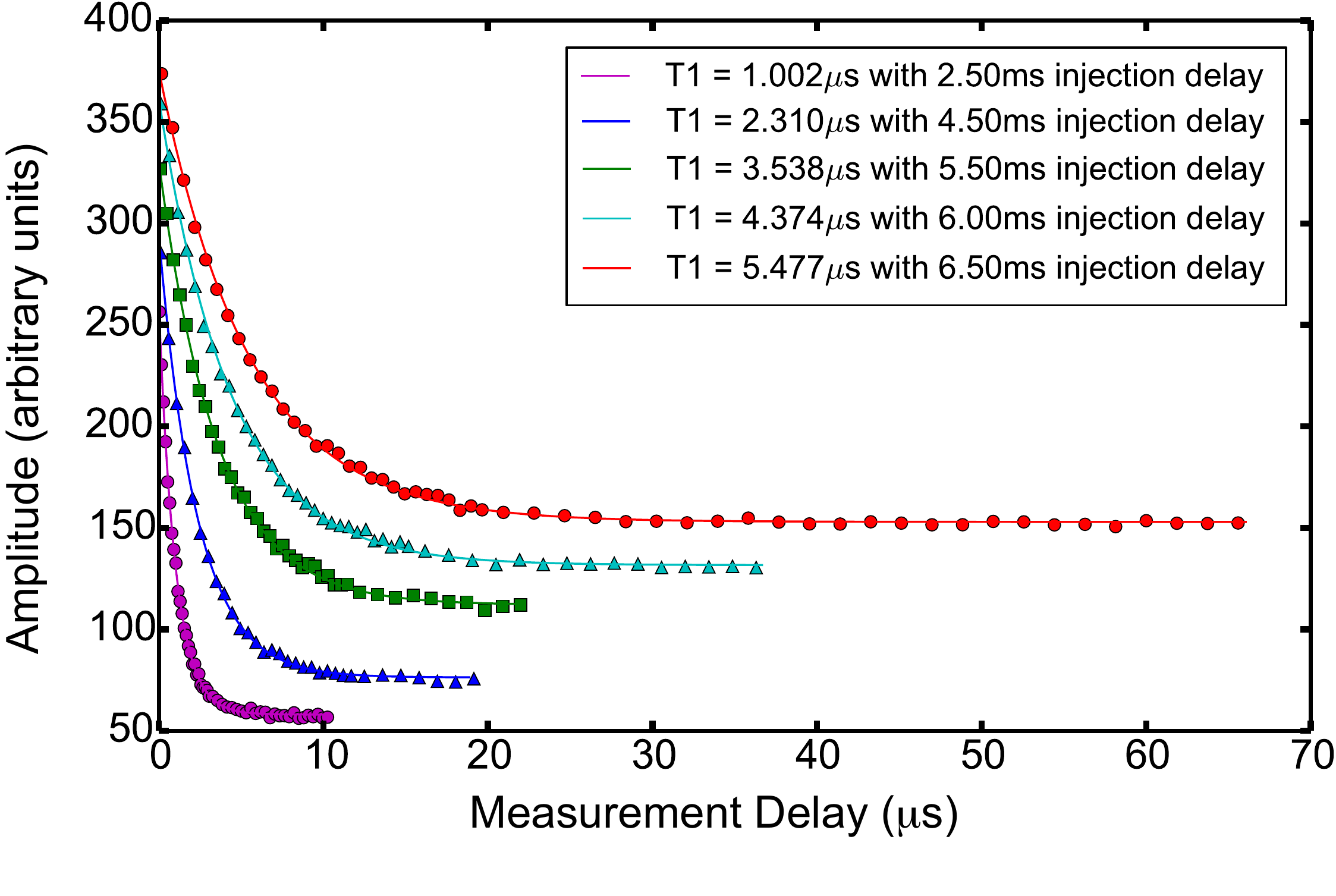}
    \caption{\textbf{Representative qubit energy decay curves.} Qubit excited state probability $P_e$ as a function of time $dt$ after a $\pi$-pulse at a few different times ($t$) after QP injection measured by high power readout (on device A3 in an aluminum cavity).  The curves are vertically offsetted for better visibility in the typical choice of $dt$'s we use for different $T_1$'s.}
    \label{T1curves}
\end{figure}

\begin{figure}[!htb]
    \centering
    \vspace{-3.2cm}
    \includegraphics[width=0.48\textwidth]{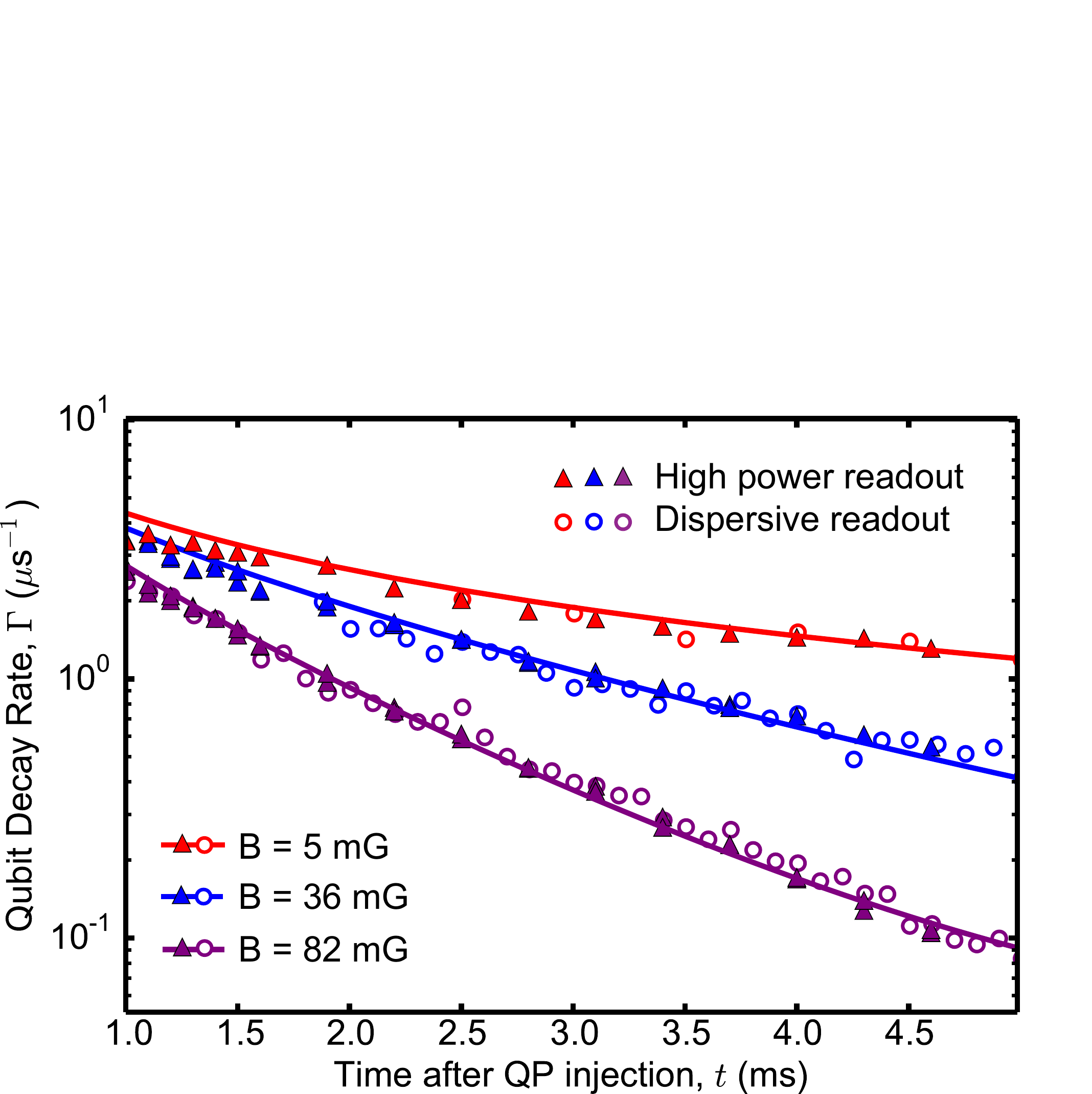}
    \caption{\textbf{Comparison of qubit decay rates measured by different readout methods.} Qubit decay rate $\Gamma$ measured by high power read-out (filled triangles) and dispersive read-out (hollow circles), plotted as a function of time $t$ after quasiparticle injection pulses for Device B1 cooled in a few selected magnetic fields $B$.  Data is shown for a limited range of $t$ for a close comparison between the two read-out methods.  Solid curves, fit to Eq.~(2) of the main text using full range of $\Gamma(t)$ data.}
    \label{HPdisp}
\end{figure}

Both readout methods measure the state of the qubit (excited state probability $P_e$) at a variable delay $dt$ ($dt\ll t$) after a $\pi$-pulse following a given delay $t$ after quasiparticle injection.  For each $t$, we then fit $P_e$ as a function of $dt$ to a single exponential function, and hence extract the qubit relaxation rate $\Gamma=1/T_1$ as a function of $t$.  Representative qubit relaxation curves are shown in Fig.~\ref{T1curves}.  For every $\Gamma(t)$ trace shown in the main text, we use high power readout in the range of 200 $\mu\mathrm{s}\leq t<8$ ms and dispersive readout for $t>1$ ms.  Data from both methods are combined together to be plotted and fit to Eq.~(2), and in overlapping region we find the $T_1$ measured by both methods agree with each other within statistical uncertainty (Fig.~\ref{HPdisp}).

For both readout methods, the measurement delays, $dt$, are chosen to be close to the estimated qubit $T_1$ under the given condition within an order of magnitude.   This is very important in a long automated measurement to maximize data-acquisition efficiency where the $T_1$ of the qubit varies by three orders of magnitude.  Such an estimate is done by an educated initial guess and active feedback on time scale of minutes.   When dispersive readout is used, because the readout signals corresponding to qubit $|g\rangle$ and $|e\rangle$ states are relatively stable, we further focus our measurement on the regime $T_1/2<dt<T_1$ where $P_e$ is maximally sensitive to the $T_1$ of the qubit (and periodically calibrate the signals for $|g\rangle$ and $|e\rangle$ states several times per hour).

When the qubit relaxation is dominated by injected quasiparticles, we find the measured $T_1$ is consistent under the same injection condition over long periods of time (\textit{i.e.}~several days).  This indicates that our injection is quite stable.  This can be seen from the inset of Fig.~3a in the main text where at small $t$ the curves for different cooling fields (measured over the course of two weeks) start from approximately the same qubit decay rate $\Gamma$.

\section{Heating from the injection pulse}

\begin{figure}[!htb]
    \centering
    \includegraphics[width=0.41\textwidth]{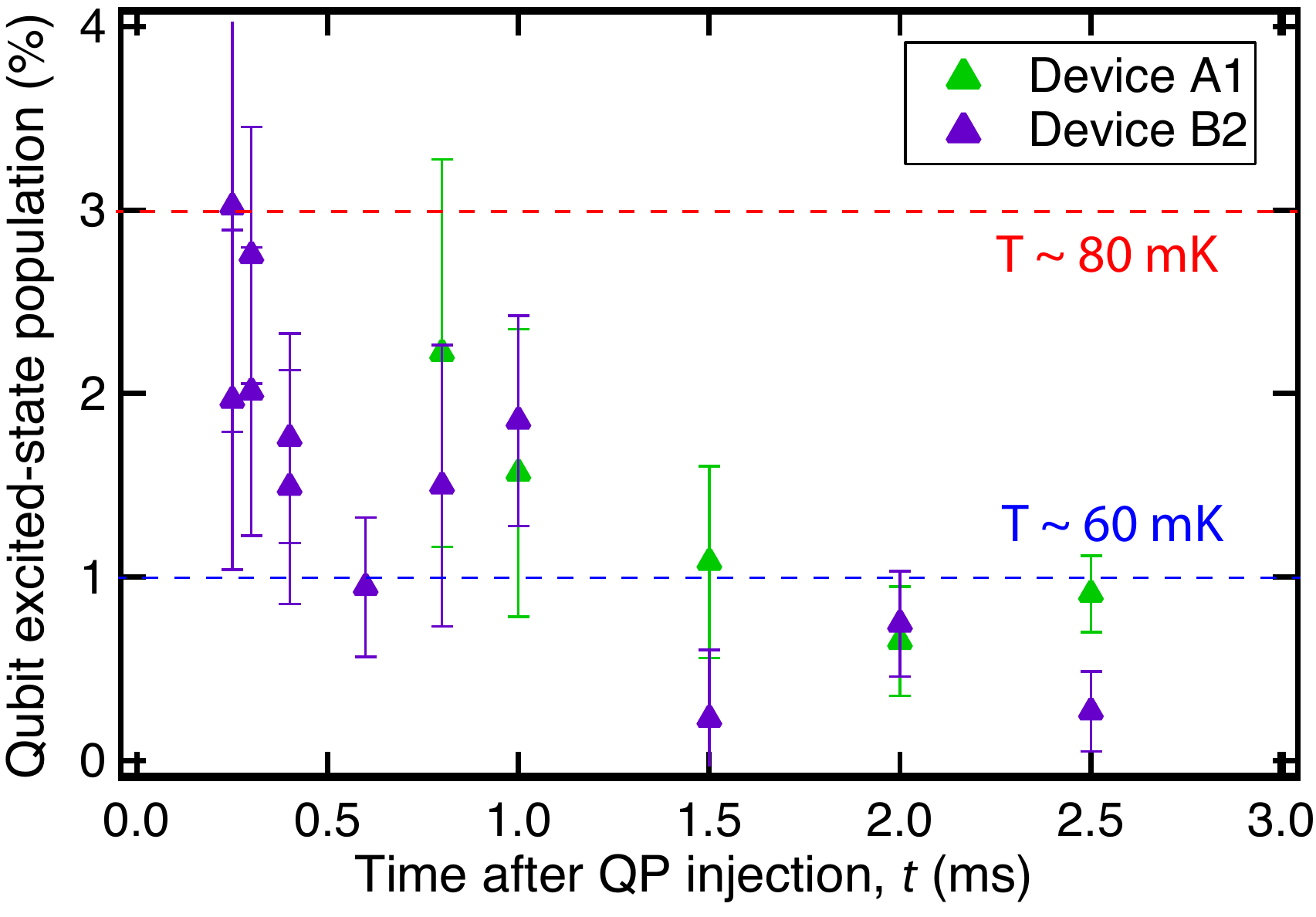}
    \caption{\textbf{Heating from the quasiparticle injection pulse.}  The $|e\rangle$-state population of the qubit is plotted as a function of time after QP injection.  The qubit effective temperatures corresponding to excited-state populations of 3\% and 1\% are about 80 mK and 60 mK respectively.  Error bars represent one s.d. statistical uncertainty in extracting the $|e\rangle$-state population from fitting the Rabi oscillations between $|e\rangle$ and the second excited state $|f\rangle$.}
    \label{population}
\end{figure}

Despite the power and length used for the QP injection pulse, we find the qubit is not heated significantly within the relevant time span of our measurements ($\gtrsim200$ $\mu$s after the injection pulse).  Since a qubit at an elevated temperature should have a relatively high probability to occupy the excited state due to thermal excitations, we can infer the temperature of the qubit at any given time after the QP injection pulse by measuring the excited state population using a protocol introduced in Ref.~\citen{Geerlings2013}.  Across different samples we find the qubit excited state population stays well below 5\% (Fig.~\ref{population}), indicating the qubit temperature does not exceed 80 mK for all time delays relevant to our measurement.  This  relatively modest heating effect is consistent with Ref.~\citen{Vool2014} where the qubit temperature after a high power microwave pulse can be more accurately measured in a fluxonium system.   The thermal population of quasiparticles at 80 mK is expected to be many orders of magnitude below the non-equilibrium $x_\qp$ observed in our device, and we further confirm that $T_1$ of our qubits shows no measurable decrease when the fridge temperature is raised from 20 mK to 100 mK as was shown in Ref.~\citen{Paik2011}.  Therefore the significantly reduced $T_1$ after the injection pulse is due to the excess amount of quasiparticles instead of an increase of temperature.

We further note that the excited state population of the qubit reflects the energy distribution of the quasiparticles when the qubit decay is dominated by quasiparticle tunneling (which is the case for majority of our measurements).  While the density of quasiparticles shortly after the injection (\textit{e.g.} $x_\qp\gtrsim10^{-4}$ at $t=200$ $\mu$s) would correspond to a thermal QP density at well over 200 mK, their energy distribution is predominantly near the gap with an effective temperature of no more than 80 mK.  We attribute this to the much shorter time scale for quasiparticles to thermalize with the phonon bath (losing their energy in excess of $\Delta$) compared with the time scale of recombination or trapping.

\section{Model for quasiparticle dynamics}

Here we consider a phenomenological model to account for diffusion, recombination, and trapping of quasiparticles. In the presence of these processes, we write the (normalized) quasiparticle density as a function of both time and spatial coordinate, $x_\qp(t,\stackrel{\scriptscriptstyle{\rightarrow}}{R})$.  Note that $x_\qp(t)$ defined in the main text is equal to $x^\JJ_\qp(t)=x_\qp(t, \stackrel{\scriptscriptstyle{\rightarrow}}{R}$ = 0) defined here if we choose the position of the Josephson junction as $\stackrel{\scriptscriptstyle{\rightarrow}}{R}$ = 0.  

The quasiparticle density obeys the following equation~\cite{eqnote}:
\be\label{diffeq}
\frac{\partial x_\qp}{\partial t} = D \nabla^2 x_\qp - r x_\qp^2 - s_0 x_\qp + g - P \sum_{i=1}^N x_\qp \, \delta \left(\stackrel{\scriptscriptstyle{\rightarrow}}{R} -\stackrel{\scriptscriptstyle{\rightarrow}}{R}_i \right)
\ee
where $D$ is the (effective) diffusion constant for quasiparticles, $r$ is the recombination coefficient, $s_0$ is the trapping coefficient which accounts for small, homogeneous (across the device) residual trapping in the absence of vortices, and $g$ is the generation rate. The factor $P$ which we call ``trapping power'' accounts for the efficiency of the quasiparticles trapping in the normal core of a vortex;
we will comment on the possible microscopic origin of $P$ in Section D below. In a thin-film device, the quasiparticles density is homogeneous across the film's thickness, so $\stackrel{\scriptscriptstyle{\rightarrow}}{R}$ is a two-dimensional vector in the plane of the device (hereinafter we assume a simplified geometry depicted in Fig.~\ref{fig:dev}), $\nabla^2$ is the Laplace operator in two dimensions, and $\stackrel{\scriptscriptstyle{\rightarrow}}{R}_i$ is the coordinate of the $i$-th vortex. The total number of vortices trapped in the device $N=N_L+N_R$ is the sum of the number of vortices in the left and right pads.

There is no simple full analytical solution of the above non-linear partial differential equation; however, considerable simplifications  are possible in experimentally relevant regimes. Below we first separately consider the cases in which vortices are absent or present in the pads, and then consider the combined effect of recombination and vortex trapping under certain approximations.

\subsection{Evolution of quasiparticle population in the absence of vortices}
\label{sec:rc}

If there are no vortices, the last term in \eref{diffeq} is absent. As quasiparticles are injected at the junction, they will initially diffuse away from the junction; then diffusion, recombination, and trapping all act to make the quasiparticle density uniform throughout the device, so at sufficiently long times we can drop the first term on the right hand side of \eref{diffeq} [see also the next section]. The remaining terms form an ordinary differential equation [also see Eq.(1) of the main text]:
\begin{equation}
	\frac{dx_\qp}{dt}=-rx^2_\qp-s_0x_\qp+g
	\label{eq:dynamics}
\end{equation}
that determines the time evolution of the quasiparticle density. The general solution of that equation can be written in the form
\be\label{xrsg}
x_\qp (t) = x_i \frac{1-r'}{e^{t/\tss}-r'} + x_0
\ee
where $x_i = x_\qp(t=0) - x_0$ is the is the normalized density of the injected quasiparticles,
\be
x_0 = \frac{\sqrt{s_0^2+4gr}-s_0}{2r}
\ee
is the steady-state density, and
\be
\frac{1}{\tss} = 2 r x_0 + s_0
\ee
is the time constant for the exponential decay at long times $t\gg \tss$. The dimensionless parameter $r'$ (with $0\le r' < 1$) quantifies the deviation from a simple exponential decay.

Since the qubit decay rate $\Gamma$ is proportional to $x_\qp$ (which is homogenous in this case), by fitting the time dependence of $\Gamma$ after quasiparticle injection [see Fig.~2 in the main text] with the function
\be
\Gamma(t) = C x_i \frac{1-r'}{e^{t/\tss}-r'} + \Gamma_0
\ee
we can estimate (or put bounds on) the parameters entering \eref{xrsg}. Then we can obtain the parameters $s$, $r$, and $g$ via the relations
\bea
r & = & \frac{r'}{(1-r')\tss x_i}\\
s_0 & = & \frac{1}{\tss}\left[1 - \frac{2 r'x_0}{(1-r')x_i}\right]  \\
g & = & \frac{x_0}{\tss}\left[1 - \frac{r' x_0}{(1-r')x_i}\right]
\eea

The recombination constant $r$ extracted from our measurements is in the range 1/(80 ns)-1/(170 ns) and agrees reasonably well with the theoretical estimate (see subsection ``Quasiparticle recombination constant") $r=21.8/(F\tau_0)$; here $\tau_0=438$ ns~\cite{kaplan1976} characterizes the electron-phonon coupling in Al, and $F$ is the phenomenological parameter associated with the phonon trapping and re-absorption in the Al film. Parameter $F$ depends~\cite{kaplan1979} on the phonon transmission coefficient $\eta$ across the Al-sapphire boundary and on the ratio of the re-absorption length $\Lambda$ to the Al film thickness $d$, and ranges from $F=1/\eta$ at $d\ll\Lambda/4$ to $F=4d/(\eta\Lambda)$ at $d\gg\Lambda/4$. The properly averaged~\cite{kaplan1979} over the transversal and longitudinal phonons value of $\eta$ yields $F\approx 5$. There is an uncertainty in the measured value of $\Lambda$, with reported values between 100 nm to 350 nm~\cite{kaplan1979}; the factor $4d/\Lambda$ possibly brings an additional factor of $\sim 2$ in $F$, resulting in $F=5-10$ quoted in the main text.

Note that in fitting $\Gamma$, only an upper bound on $x_0$ can be placed, $x_0 \le \Gamma_0/C$, since $\Gamma_0$ can receive contributions from other (non-quasiparticle) decay mechanisms.  Also taking into account $x_0\geq0$, we obtain a bound for $s_0$:
\be
\frac{1}{\tss}\left[1 - \frac{2 r' \Gamma_0}{(1-r')x_i C}\right] \leq s_0 \leq \frac{1}{\tss}
\ee
For most of our experimental conditions, we find the bounded range for s is comparable or smaller than statistical uncertainties, therefore giving a relatively accurate determination of $s_0$.  We also obtain an upper bound\cite{note_g} for $g$:
\be
g \le \frac{(1-r')x_i}{4 r' \tss} = \frac{1}{4 r \tss^2}
\ee
Using the parameters extracted for Device B1, the latter expression gives, $g < 2\times 10^{-4}$s$^{-1}$, in excellent agreement with the value estimated using Eq.~(3) in the main text [see also Fig.~4c and 4d], as well as with that recently estimated in an experiment with a fluxonium qubit.\cite{Vool2014}

\subsection{Trapping of quasiparticles by vortices}
\label{sec:tp}

\begin{figure*}[hbt]
\vspace{0.5cm}
\begin{center}
    \includegraphics[width=0.92\textwidth]{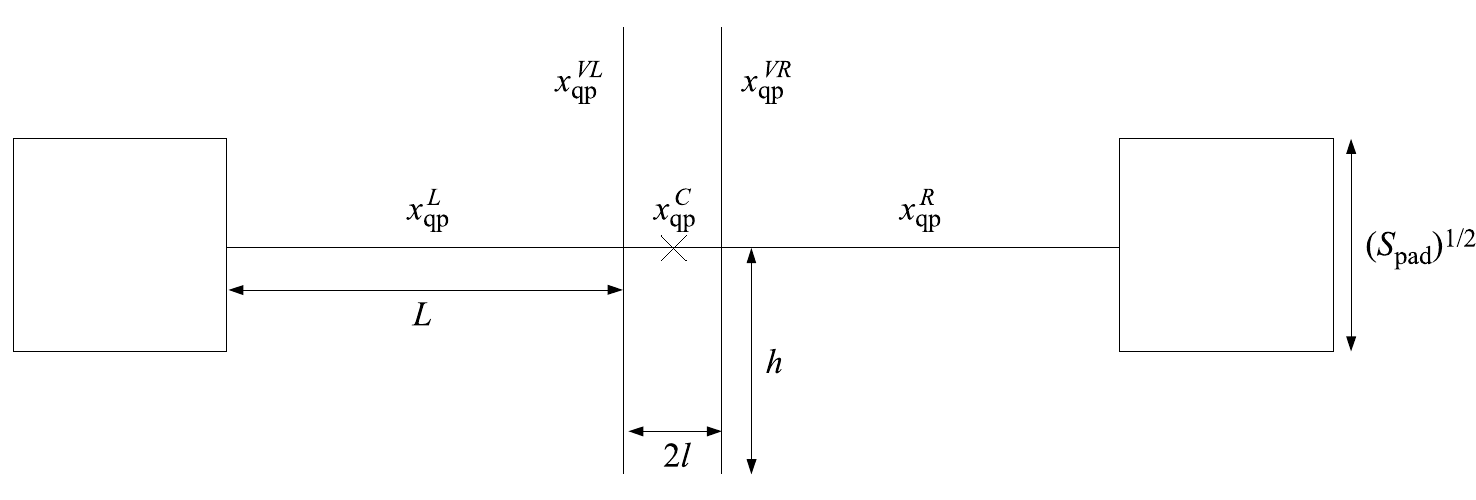}
\end{center}
\caption{\textbf{Simplified geometry used to model the Type B devices.} The thin lines represent wires of width $W$ much smaller than their lengths. See Fig.~\ref{fig:cap} and the text after \eref{l0lkeq} for a more realistic model of the vertical capacitor pads.}
\label{fig:dev}
\end{figure*}

When the qubit is cooled in a sufficiently high magnetic field ($\gtrsim 10$~mG for Type B devices), superconducting vortices are trapped in the pads. The localization of the vortices far from the junction means that at all times the quasiparticle density is not uniform, so that in general the diffusion term in \eref{diffeq} cannot be neglected. On the other hand, if quasiparticle trapping is the most efficient mechanism for relaxing the quasiparticle density, we can neglect the small effects of generation and recombination and reduce \eref{diffeq} to
\be\label{diffeq2}
\frac{\partial x_\qp}{\partial t} = D \nabla^2 x_\qp - s_0 x_\qp - P \sum_{i=1}^N x_\qp \, \delta \left(\stackrel{\scriptscriptstyle{\rightarrow}}{R} -\stackrel{\scriptscriptstyle{\rightarrow}}{R}_i \right)\,
\ee
(The injection of quasiparticles is introduced by a time-dependent boundary condition, similar to Ref.~\citen{Rajauria2009}.)

We can solve \eref{diffeq2} based on a geometric model of the superconducting electrodes shown in Fig.~\ref{fig:dev} consisting of pads, gap capacitors and connecting wires.  This level of complexity is required in order to properly account for the time scale of diffusion when there are unequal number of vortices in the two pads, and the resultant calculation is fairly complicated.  The goal of the following calculation is to solve the spatial mode with the slowest decay rate $s$, that is a certain spatial distribution of $x_\qp$ that decays synchronously across the device, \textit{i.e.}~$x_\qp(\stackrel{\scriptscriptstyle{\rightarrow}}{R}, t)=x_\qp(\stackrel{\scriptscriptstyle{\rightarrow}}{R}, t=0)e^{-st}$.  One instructive limiting case to see the effect of vortices is when the diffusion time ($\sim L^2/D$) is much shorter than the decay time ($1/s$) so that the spatial gradient of $x_\qp$ in \eref{diffeq2} can be neglected.  Then by integrating \eref{diffeq2} over the area $A$ of the device one can immediately see: $sA=s_0A+NP$.  Therefore the decay rate multiplied by the device area is the macroscopic observable directly linked to the trapping power we defined.  More generally, we find for any given number of vortices $N_L$ and $N_R$ on the two pads, this decay rate $s$ can be numerically computed from Eqs.~\rref{l0lkeq}-\rref{tqp} derived below and compared with our measured data as shown in Fig.~\ref{fig:steps}.

Our geometric model of the electrodes is composed of two square pads of area $S_\mathrm{pad}$, each connected to the gap capacitor by a narrow wire of length $L$ and width $W$, with $W\ll L$. Due to their narrow widths, we treat each wire as one-dimensional. The quasiparticle density, normalized by the density of Cooper pairs,~\cite{qp_prb} is denoted as $x_{\qp}^L(t,y)$ [$x_{\qp}^R(t,y)$] in the wire to the left [right] of the capacitor; the corresponding coordinate $y$ runs from $-L$ to $0$ [$0$ to $L$].
As a simple model for the gap capacitor, we use a thin wire (width $W$) extended
to length $h$ on both sides of the central wire; below we will discuss a more realistic model for the gap capacitor. We assume symmetry with respect to the horizontal line connecting the centers of the pads; then we only need to introduce two densities for the left and right upper halves of the capacitor, denoted by $x_\qp^{VL}(t,y)$ and $x_\qp^{VR}(t,y)$, respectively. Here coordinate $y$ runs from $0$ at the top to
$h$ at the junction with the horizontal wire. The two plates of the gap capacitors are at a distance $2l$ from each other, with the junction placed
at the center of a short and thin wire ($l \ll h$, $L$). The quasiparticle density in this central wire is denoted by $x_\qp^C(t,y)$, with the spatial coordinate $y \in [-l,l]$.

Since the qubit is affected only by quasiparticles in the junction vicinity, we are interested in the time evolution of $x_\qp^C(t,0)$. To find the decay rate of $x_\qp^C(t,0)$, we start by noting that except in the pads, the quasiparticle densities obey a simple diffusion equation [cf. \eref{diffeq2}]
\be\label{dew}
\frac{\partial x_\qp^i}{\partial t} = D \frac{\partial^2 x_\qp^i}{\partial y^2} - s_0 x_\qp^i
\ee
with $i=L$, $R$, $C$, $VL$, or $VR$. The general solution to this equation can be written in the form
\be
x_\qp^i (t,y) = e^{-s t}\left[\alpha^i \cos ky + \beta^i \sin ky \right]\, ,
\ee
where $s$ and $k$ are related by
\be\label{lambda}
s = Dk^2 + s_0 \, .
\ee
Here we are interested in the spatial mode with the lowest decay rate; it is obtained by finding the smallest $k$ that satisfies the boundary conditions which we now discuss.

First, we require that no current leaves the top (and bottom, by symmetry) of the vertical wires:
\be
\frac{\partial x_\qp^{Vj}}{\partial y}\Big|_{y=0} = 0
\ee
where, from now on, $j=L$ or $R$.
These conditions imply $\beta^{Vj}=0$. Next, we require continuity of density at the two cross points:
\be\label{dcc}
x_\qp^C(\pm l) = x_\qp^j(0) =  x_\qp^{Vj}(h) \, ,
\ee
where in the first term the positive sign is to be used for $j=R$ and the negative one for $j=L$.
From the last equality we find immediately
\be
\alpha^{Vj} = \alpha^j / \cos kh \, .
\ee
Taking the sum and difference of the first equality in \eref{dcc} with $j=R$ and $j=L$ gives
\be
\alpha^C = \left(\alpha^R + \alpha^L \right)/2 \cos kl \, , \quad \beta^C = \left(\alpha^R - \alpha^L \right)/2 \sin kl \, .
\ee
We then require current conservation at the two cross points:
\be
\frac{\partial x_\qp^{L}}{\partial y}\Big|_{y=0} + 2 \frac{\partial x_\qp^{VL}}{\partial y}\Big|_{y=h} = \frac{\partial x_\qp^{C}}{\partial y}\Big|_{y=-l}
\ee
for the left-side cross point and
\be
\frac{\partial x_\qp^{C}}{\partial y}\Big|_{y=l} + 2 \frac{\partial x_\qp^{VR}}{\partial y}\Big|_{y=h} = \frac{\partial x_\qp^{R}}{\partial y}\Big|_{y=0}
\ee
for the right one. Using the results above, these conditions can be written explicitly as
\be
\beta^L - 2\alpha^L \tan kh = \frac{\alpha^R+\alpha^L}{2} \tan kl +\frac{\alpha^R - \alpha^L}{2} \cot kl \,
\ee
\be
-\frac{\alpha^R+\alpha^L}{2} \tan kl +\frac{\alpha^R - \alpha^L}{2} \cot kl -2\alpha^R \tan kh = \beta^R
\ee
and enable us to express $\beta^{L,R}$ in terms of $\alpha^{L,R}$.

We note that the boundary conditions discussed so far do not involve the pads, and let us write the densities in the left and right wires in the explicit forms
\bea
x_\qp^L &=& e^{-s t}\bigg[\alpha^L \cos ky +\bigg(\frac{\alpha^R+\alpha^L}{2} \tan kl  \label{xl} \\
&& +\frac{\alpha^R - \alpha^L}{2} \cot kl + 2\alpha^L \tan kh\bigg) \sin ky\bigg] \nonumber\\
x_\qp^R &=& e^{-s t}\bigg[\alpha^R \cos ky +\bigg(-\frac{\alpha^R+\alpha^L}{2} \tan kl  \label{xr} \\
&& +\frac{\alpha^R - \alpha^L}{2} \cot kl - 2\alpha^R \tan kh\bigg) \sin ky\bigg] \nonumber
\eea
where only the two coefficients $\alpha^L$ and $\alpha^R$ are yet to be determined.
In the pads, we assume the diffusion time $S_\mathrm{pad}/D$ to be the shortest time scale (in particular, shorter than the inverse of the average trapping rate $N_{L,R} P/S_\mathrm{pad}$), so that the densities in the pads, $x_\qp^{Pj}$, can be taken as uniform;
the value of trapping power $P$ we ultimately extract from the experiment justify this assumption. By integrating \eref{diffeq} over, \textit{e.g.}, the right pad, we find
\be
S_\mathrm{pad} \frac{\partial x_\qp^{PR}}{\partial t} = I_W - S_\mathrm{pad} s\, x_\qp^{PR} - N_R P \, x_\qp^{PR} \, ,
\ee
where $I_W$ represent the quasiparticle current going from the pad into the wire. By current conservation we also have
\be
I_W = - W\, D \frac{\partial x_\qp^R}{\partial y}\Big|_{y=L} \, ,
\ee
while continuity requires $x_\qp^{PR} = x_\qp^R|_{y=L}$.
Therefore, we arrive at the following boundary condition at position $y=L$:
\be\label{rpbc}
\frac{\partial x_\qp^R}{\partial y} = -\frac{1}{W D} \left.\left[S_\mathrm{pad} \left(\frac{\partial x^R_\qp}{\partial t}+s x_\qp^R\right) +  N_R P \, x_\qp^R \right] \right|_{y=L}.
\ee
Similarly, at the boundary with the left pad ($y=-L$) we have:
\be\label{lpbc}
\frac{\partial x_\qp^L}{\partial y} = \frac{1}{W D} \left.\left[S_\mathrm{pad} \left(\frac{\partial x_\qp^L}{\partial t}+s x_\qp^L\right) +  N_L P \, x_\qp^L \right] \right|_{y=-L}.
\ee
Substituting \esref{xl} and \rref{xr} into \esref{rpbc}-\rref{lpbc}, and requiring the existence of a non-trivial solution, leads to an equation for $k$. Introducing
dimensionless variable $z =kL$, this equation is:
\begin{widetext}
\be\label{fullkeq}\begin{split}
& \left[z \left(\tan z + f(z)\right) + \left(a z^2 -\bar{N} \frac{P\tau_D}{A_W}\right)\left(1 - \tan z f(z) \right) \right]^2 -
\left(\frac{\Delta N}{2} \frac{P\tau_D}{A_W}\right)^2\left(1 - \tan z f(z) \right)^2 \\
& -\frac14 \left[\tan \left(z\frac{l}{L}\right) +\cot \left(z\frac{l}{L}\right) \right]^2
\left\{\left[z -\tan z\left(a z^2 -\bar{N} \frac{P\tau_D}{A_W}\right) \right]^2-\left(\frac{\Delta N}{2} \frac{P\tau_D}{A_W}\right)^2 \tan^2 z\right\} = 0
\end{split}\ee
\end{widetext}
with $\tau_D=L^2/D$ the diffusion time along the left and right wires, $A_W=LW$ the area of the wires (from a cross point to a pad), $a=S_\mathrm{pad}/A_W$, $\bar{N} = (N_L+N_R)/2$ the average number of vortices, $\Delta N = N_R -N_L$, and
\be
f(z) = \frac12 \tan\left(z \frac{l}{L}\right) - \frac12 \cot\left(z \frac{l}{L}\right) + 2 \tan \left(z \frac{h}{L}\right).
\ee

We can simplify \eref{fullkeq} as follows: even at large number of vortices, the variable $k$ being as small as possible means that it is at most of order $1/L$, \textit{i.e.}, $z$ is at most of order 1. Then we note that for the short wire inside the gap capacitor we have $l \ll L$ and thus, keeping only the leading term, in the limit $l\to 0$ we find
\be\label{l0lkeq}\begin{split}
&\left[z-\tan z \left(a z^2 -\bar{N} \frac{P\tau_D}{A_W}\right)\right]\bigg[z\left(\tan z + 2 \tan\left(z \frac{h}{L}\right) \right)\\ &+\left(a z^2 -\bar{N} \frac{P\tau_D}{A_W}\right)\left(1-2 \tan \left(z \frac{h}{L}\right)\tan z\right)\bigg] \\ &+\left(\frac{\Delta N}{2} \frac{P\tau_D}{A_W}\right)^2 \tan z \left(1-2 \tan \left(z \frac{h}{L}\right)\tan z\right) = 0.
\end{split}\ee

\begin{figure}[tbh]
\begin{center}
    \includegraphics[width=0.26\textwidth]{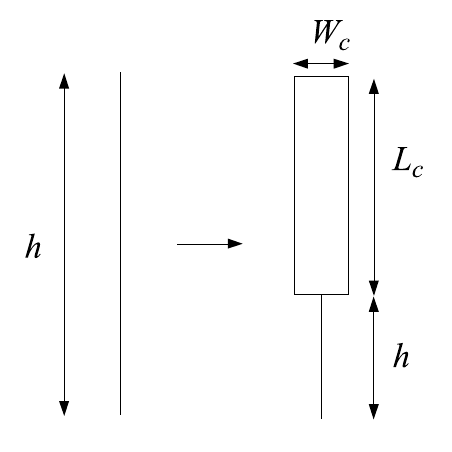}
\end{center}
\caption{\textbf{Improved model for the capacitor of the Type B devices.}  To the the thin wire of width $W$ and length $h$ on the left, we add a wider wire of width $W_c$ and length $L_c$ to form the structure on the right.}
\label{fig:cap}
\end{figure}

\begin{figure}[tbp]
\vspace{0.5cm}
\begin{center}
   \includegraphics[width=0.48\textwidth]{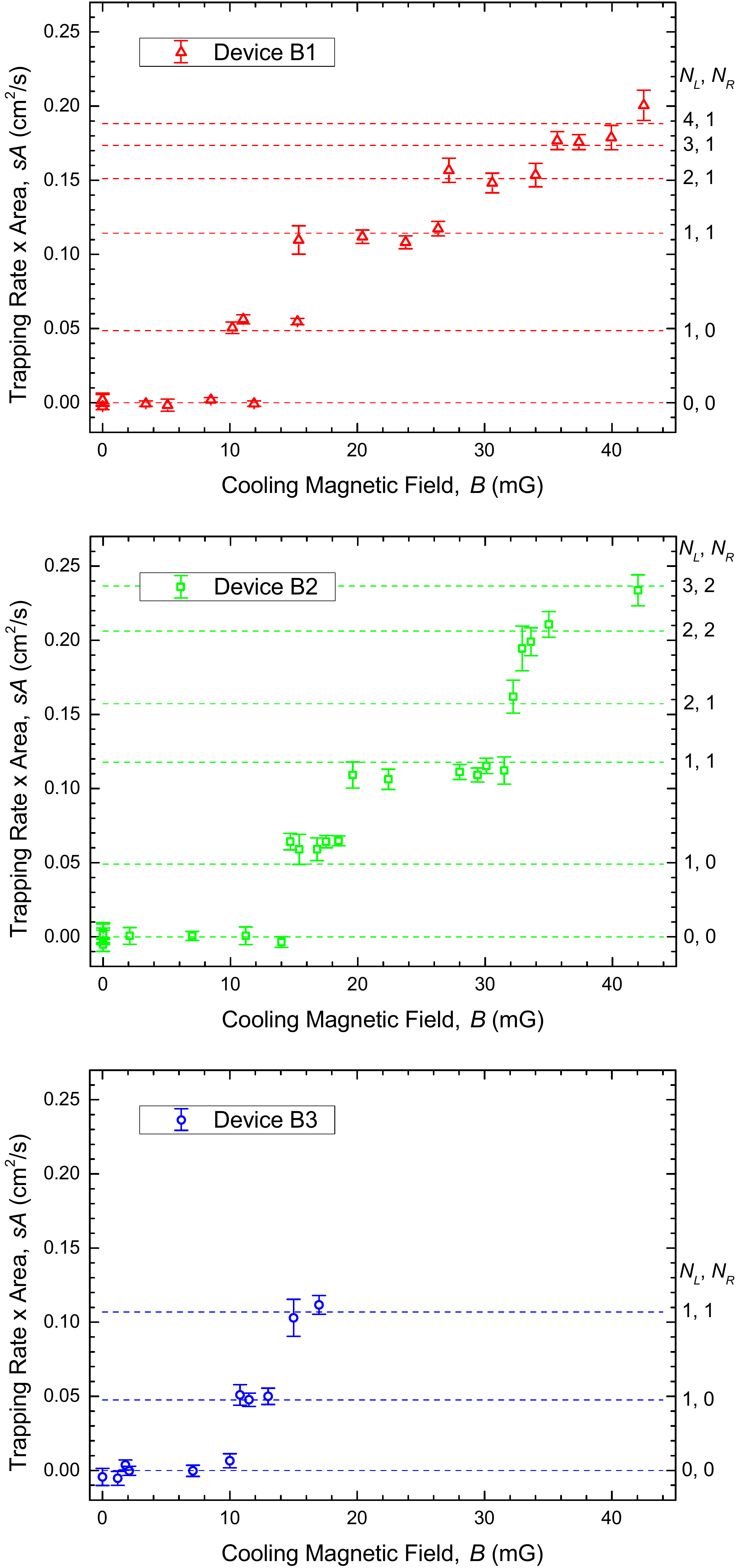}
\end{center}
\caption{\textbf{Fitting the discrete steps in quasiparticle trapping rates}.  Normalized trapping rates for (top to bottom) devices B1, B2, and B3 as a function of cooling magnetic field, with error bars showing the range of fluctuations for repetitive measurements over each thermal cycle. Horizontal dashed lines: rates calculated as explained in the text for the number of vortices in the two pads listed by the two numbers to the right of each line.  \textit{e.g.}~21 represents $N_L=2$, $N_R=1$ or vice versa.}
\label{fig:steps}
\end{figure}

As an improvement to our description of the device, we further model the gap capacitor by adding to the thin wires of width $W$ and length $h$, wider sections of width $W_c$ and length $L_c$, as shown in Fig.~\ref{fig:cap}.

Requiring again current conservation and continuity of the density, we find that this more realistic geometry can be incorporated into the formulas above by the substitution
\be\label{capsub}\begin{split}
\tan \left(z \frac{h}{L}\right)  \to \frac{\cos \left(z \frac{L_c}{L}\right) \sin \left(z \frac{h}{L}\right)+ \frac{W_c}{W} \sin \left(z \frac{L_c}{L}\right) \cos \left(z \frac{h}{L}\right)}{\cos \left(z \frac{L_c}{L}\right) \cos \left(z \frac{h}{L}\right)- \frac{W_c}{W} \sin \left(z \frac{L_c}{L}\right) \sin \left(z \frac{h}{L}\right)}.
\end{split}\ee

Equation \rref{l0lkeq} with the substitution in \eref{capsub} makes it possible to calculate the expected decay rates $1/\tss$ of the quasiparticle density based on the geometry of each device. Indeed, given the values of trapping power $P$ and diffusion constant $D$ (which determines the diffusion time $\tau_D$), and the number of vortices in each pad, we can solve \esref{l0lkeq}-\rref{capsub} for $z$ and then find $s$ from
\be\label{tqp}
s = \frac{1}{\tau_D} z^2 + s_0 \, .
\ee

Equations \rref{l0lkeq}-\rref{capsub} can be solved analytically only in limiting cases. For example, in the case of small trapping power $P\to 0$, the leading order result for the decay rate is
\be\label{slf}
s \simeq \left(N_L+N_R\right) \frac{P}{A} + s_0
\ee
where $A$ is the total area of the device. This expression explains why, after subtracting the device-dependent homogeneous trapping coefficient $s_0$ and multiplying by the total device area, the decay rates for the different devices are approximately equal and quantized. In the opposite limit of large trapping power (and/or large number of vortices), we find that when the area of the capacitor $A_c = 2(L_c W_c + h L)$ is small compared to that of the central wire, $ A_c\ll A_W$, the variable $z$ to be substituted in \eref{tqp} is given by
\be\label{zhf}
z \simeq \frac{\pi}{2} \frac{1}{1+A_c/A_W} \, .
\ee
This formula qualitatively explains why in Fig.~3b of the main text the decay rate of Device B2 (with a smaller gap capacitor) saturates at a higher value than the decay rate of Device B1 (with a bigger gap capacitor).

For a quantitative comparison with experiments, we numerically solve \esref{l0lkeq}-\rref{capsub} to find $z$, and then substitute this number into \eref{tqp}. In Fig.~\ref{fig:steps} we show with dashed lines the results of such calculations with $P=0.067$~cm$^2$s$^{-1}$, $D=18$~cm$^2$s$^{-1}$, and the number of vortices in each pad given to the right of each panel; the homogeneous trapping coefficient $s_0$ has been subtracted out to facilitate comparison among the devices.  The results calculated with the same $P$ and $D$ at higher magnetic field has been plotted as the dashed line in Fig.~3b of the main text, where $N_L$ and $N_R$ are assumed to be equal and increasing linearly with magnetic field.  The diffusion constant $D$ is determined by the saturation level of $s$ and the geometry of the device [see Eqs.~(\ref{tqp}) and (\ref{zhf}) and $\tau_D=L^2/D$], and its uncertainty is determined to be $\sim10\%$ (or $\pm$ 2 cm$^2$s$^{-1}$) based on the statistical fluctuation of $s$ (Fig.~3b of the main text).  (Note that this does not include possible error due to the simplified model of the actual device geometry, which by our estimate should not increase the overall uncertainty significantly.)  The trapping power $P$ is mostly determined by the height of the first 2-3 steps in $sA$, slightly adjusted for the finite speed of diffusion [see Eq.~(\ref{slf})].  We determine $\sim7\%$ (or $\pm$ 0.005 cm$^2$s$^{-1}$) uncertainty on $P$ based on the statistical fluctuation of $s$ for the $(N_L, N_R)=(1,0)$ and $(1,1)$ states ($\sim7\%$) and uncertainty propagated from $D$ ($\sim 2\%$).

We note that the agreement between the predicted steps in $sA$ and the measured data (Fig.~\ref{fig:steps}) could be further improved by considering vortices entering the device two at a time (using the vortex number series $(N_L, N_R)=(0,0), (1,1), (2,2),$ etc.~for the first few steps).  Under this interpretation the trapping power for a single vortex would be a factor of two smaller, \textit{i.e.}~0.034 cm$^2$s$^{-1}$, since each step change in $sA$ corresponds to the entry of two vortices.  While we cannot entirely exclude this possibility, we argue it is an unlikely scenario since such a high degree of symmetry seems unwarranted.  The actual critical field $B_k$ for the two trapping pads in each device should vary from each other to a similar degree as they vary from device to device, typically a few mG, which is larger than the $\lesssim1$ mG increments we typically use in changing cooling magnetic field near $B_k$.  Furthermore, variation of a few mG in $B_k$ is consistent with the width of the first step in $sA$ (labeled ``10" in Fig.~\ref{fig:steps}).  The second step (labeled ``11" in Fig.~\ref{fig:steps}) being much wider than the first step also suggests a vortex number series of $(N_L, N_R)=(0,0), (1,0), (1,1), (2,1)$, etc.~(plotted in Fig.~\ref{fig:steps} and assumed in the main text) as the more plausible scenario.  

Without changing the assumption on vortex numbers, the agreement between theory and experiment could also be improved,  especially for Device B2, by introducing a non-homogenous background trapping rate concentrated in one of the pads instead of using a homogenous trapping coefficient $s_0$.  The origin of the background trapping rate remains to be studied in the future.

\subsection{Evolution of quasiparticle population in the presence of vortex trapping and recombination}

Now that we have analyzed the quasiparticle population dynamics in the presence of either recombination or trapping from vortices, here we attempt to approximately combine both results.

In the previous section we have solved the slowest linear decay mode in the presence of vortices (neglecting quasiparticle recombination and generation) so that $x_\qp(\stackrel{\scriptscriptstyle{\rightarrow}}{R}, t)=x^\JJ_\qp e^{-st}\psi(\stackrel{\scriptscriptstyle{\rightarrow}}{R})$, where $x^\JJ_\qp=x_\qp(\stackrel{\scriptscriptstyle{\rightarrow}}{R}$ = 0) is the normalized quasiparticle density near the Josephson junction (simply quoted as $x_\qp$ in the main text), $\psi(\stackrel{\scriptscriptstyle{\rightarrow}}{R})$ describes the spatial mode (normalized to $\psi(\stackrel{\scriptscriptstyle{\rightarrow}}{R}$ $= 0 ) = 1$) satisfying
\begin{equation}\label{eq:substitution}
D \nabla^2 \psi(\stackrel{\scriptscriptstyle{\rightarrow}}{R}) - s_0 \psi(\stackrel{\scriptscriptstyle{\rightarrow}}{R}) - P \sum_{i=1}^N \psi(\stackrel{\scriptscriptstyle{\rightarrow}}{R})\,\delta \left(\stackrel{\scriptscriptstyle{\rightarrow}}{R} -\stackrel{\scriptscriptstyle{\rightarrow}}{R}_i \right)=-s\psi(\stackrel{\scriptscriptstyle{\rightarrow}}{R})
\end{equation}
where $s$ is the decay rate of the mode obtained from Eqs.~\rref{l0lkeq}-\rref{tqp}.

Now we add recombination and generation to the dynamics assuming they are relatively weak, so the spatial mode is not altered significantly and $x_\qp$ can still be factorized as $x_\qp(\stackrel{\scriptscriptstyle{\rightarrow}}{R}, t)=x^\JJ_\qp(t)\psi(\stackrel{\scriptscriptstyle{\rightarrow}}{R})$.  Under this approximation we can use the linear decay term with rate $s$ [the RHS of the Eq.~(\ref{eq:substitution})] to replace the diffusion and trapping terms [the LHS of the Eq.~(\ref{eq:substitution})] in Eq.~(\ref{diffeq}) and get:
\begin{equation}
	\frac{dx^\JJ_\qp}{dt}=-r(x^\JJ_\qp)^2-sx^\JJ_\qp+g
	\label{eq:dynamics}
\end{equation}
This is the ordinary differential equation for quasiparticle dynamics we have used in the main text [Eq.~(1)].  We can then follow the same procedures in Section A to derive the relations between fit parameters of Eq.~(2) of the main text and $r$, $s$ and $g$, except that $s_0$ in Section A should be replaced by $s$.  Therefore the trapping rates $s$ extracted in the main text based on this equation in the presence of vortices corresponds to the linear decay rate solved in Section B.

\subsection{Microscopic model for trapping power}
\label{sec:microscopic}
In modelling QP trapping by a vortex, we treated the latter as a point-like object, see \eref{diffeq}. In a more realistic (albeit still crude) model, we can expect trapping of quasiparticles to be effective over a finite region around the vortex core, and the trapping rate to be related to the quasiparticle relaxation by electron-electron (ee) and electron-phonon (ep) interactions. This can be taken into account via the replacement
\be\label{pm}
P\delta (\stackrel{\scriptscriptstyle{\rightarrow}}{R} -\stackrel{\scriptscriptstyle{\rightarrow}}{R}_i) \to \frac{1}{\tau_{\rm n}} \theta \left(R_{\rm c} - \left|\stackrel{\scriptscriptstyle{\rightarrow}}{R} -\stackrel{\scriptscriptstyle{\rightarrow}}{R}_i\right|\right)
\ee
in \eref{diffeq}. Here $R_{\rm c}$ is the radius of the vortex core modeled as a normal-state disk ($R_{\rm c}$ is of the order of the coherence length $\xi$), $\theta(R)$ is the Heaviside step function, and $1/\tau_{\rm n}=1/\tau_{\rm ee}+1/\tau_{\rm ep}$ is the microscopic electron relaxation rate. After integrating \eref{pm} over the pad area, we can relate the trapping power $P$ to these parameters:
\be
P = \frac{1}{\tau_{\rm n}} \pi R^2_{\rm c}\,.
\label{Pest}
\ee
Estimates of the two contributions to $1/\tau_{\rm n}$ can be performed using the theory of electron-electron interactions in disordered conductors~\cite{aa_review} and the conventional theory of electron-phonon interaction in metals.\cite{abrikosov} The material-specific input parameters (electron density of states and the deformation potential) for Al are known from literature, see, {\sl e.g.}, Ref.~\onlinecite{kaplan1976}, and we know the parameters specific for our devices (the film thickness and diffusion constant) from the performed measurements.  We find that the electron-electron and electron-phonon mechanisms contribute comparably to $1/\tau_{\rm n}$ at a level $1/\tau_{\rm ee}\approx 6\times10^6\,{\rm s}^{-1}$, although it should be stressed that these rates are not accurately known~\cite{Ullom2000}. There is also some uncertainty about the coherence length $\xi$, but the ``dirty-limit'' estimate $\xi\approx 100$~nm seems reasonable for a $\sim 80$~nm-thick film. Evaluation of $P$ using \eref{Pest} and assuming $R_\mathrm{c}=\xi$ yields $P\approx 0.004$ cm$^2$s$^{-1}$, about 18 times lower than found in the experiment. However, the assumption $R_\mathrm{c} = \xi$ may be underestimating the effective area where quasiparticles can be trapped. Indeed, a treatment of gap suppression near a vortex\cite{Golubov1993} gives $R_\mathrm{c} \approx 2.7\xi \approx 270$~nm and using this estimate [also employed in a recent work, Ref.~\onlinecite{Nsanzineza2014}] we find $P$ within a factor of 3 of the measured value. Also, it is not clear whether the different group velocities of quasiparticles in the superconductor and in the normal core of the vortex should affect the effective quasiparticle trapping rate\cite{ullom}. Given the simplified nature of the model and the uncertainties about the values of parameters entering it, the disagreement between measured and estimated trapping powers by a factor of 3 to 18 is not surprising.

Recently, Nsanzineza and Plourde~\cite{Nsanzineza2014} used a similar phenomenological model of quasiparticle trapping from vortices to interpret an observed increase in quality factor of a superconducting resonator as a function of magnetic field.  While Ref.~\citen{Nsanzineza2014} suggested their data is consistent with a microscopic electron relaxation rate of about $3\times10^6$s$^{-1}$, (similar to our estimate above,) we note the experiment can be better interpreted in terms of trapping power so that it is independent of microscopic assumptions.  The implied trapping power $P\approx0.024$ cm$^2$s$^{-1}$ is within a factor of 3 of our measured value, which is a decent agreement given that Ref.~\citen{Nsanzineza2014} does not have a direct probe of quasiparticle dynamics, and the extracted $P$ (or $\tau_{\rm n}$) inevitably depends on a list of unknown parameters such as recombination rate, diffusion constant, generation rate, kinetic inductance fraction, etc.  The quantization of trapping rates we have observed is a much more direct and accurate measurement of the quasiparticle trapping effect of a vortex, and should facilitate development of more quantitative microscopic theories addressing the vortex-quasiparticle interaction in superconductors. 

\subsection{Quasiparticle density distribution in the presence of vortices}

Although the microscopic mechanisms remain to be further explored, the measured trapping power $P$ gives us a quantitative understanding of the overall strength of the vortex-quasiparticle interaction.  One important conclusion we can draw from our measured ratio of $P/D\approx4\times10^{-3}\ll 1$ (based on $P=0.067$ cm$^2$s$^{-1}$, $D=18$ cm$^2$s$^{-1}$) is that the spatial distribution of QP density in a 2D extended film is close to homogenous even in the presence of vortices.  This approximate homogeneity has been assumed for the quasiparticle density in each pad, $x^{Pj}_\qp$, in subsection VI-B, and here we provide a justification.  

We first consider one vortex at $\stackrel{\scriptscriptstyle{\rightarrow}}{R}$ $=0$ in a film that is extended in 2D for $|\stackrel{\scriptscriptstyle{\rightarrow}}{R}|<r$ with no other quasiparticle loss mechanisms within this area.  We assume there is no QP generation source in the entire system, so there exists a slowest decay mode so that $x_\qp(\stackrel{\scriptscriptstyle{\rightarrow}}{R}, t) = x_\qp(\stackrel{\scriptscriptstyle{\rightarrow}}{R}) e^{-s t}$, and
\begin{equation}\label{difftrapping}
	D\nabla^2 x_\qp(\stackrel{\scriptscriptstyle{\rightarrow}}{R}) - \frac{P}{\pi R^2_c}x_\qp(\stackrel{\scriptscriptstyle{\rightarrow}}{R})\,\theta(R_{c}-|\stackrel{\scriptscriptstyle{\rightarrow}}{R}|)= - s x_\qp(\stackrel{\scriptscriptstyle{\rightarrow}}{R})
\end{equation}
Here for simplicity we use a step function to model the trapping power as was done in the previous subsection, but the subsequent conclusion is qualitatively independent of this assumption.  The value of $s$ depends on the geometry and loss mechanisms\cite{QPdist_note} in the region of $|\stackrel{\scriptscriptstyle{\rightarrow}}{R}|>r$, but our analysis only relies on $s\geq0$ which does not require knowledge of the $|\stackrel{\scriptscriptstyle{\rightarrow}}{R}|>r$ region.  Equation (\ref{difftrapping}) can be simplified in cylindrical coordinates ($\rho, \theta$), considering rotational symmetry when $\rho\ll r$, $x_\qp(\stackrel{\scriptscriptstyle{\rightarrow}}{R})=x_\qp(\rho)$,
\begin{equation}\label{cyltrapping}
	D\frac{1}{\rho}\frac{d}{d\rho}\left[\rho\frac{dx_\qp}{d\rho}\right] = \frac{P}{\pi R^2_c}x_\qp \,\theta(R_c-\rho) - s x_\qp
\end{equation}
Noting $s\ll P/(\pi R^2_c)$, we solve \eref{cyltrapping} to first order in $P/D$ for $\rho\leq R_c$:
\begin{equation}
	\frac{x_\qp}{x^0_\qp} \approx 1 + \frac{P}{4\pi D}\left(\frac{\rho}{R_c}\right)^2 < 1.001
\end{equation}
where $x^0_\qp = x_\qp(\rho=0)$ is the QP density at the center of the vortex.  Noting $s\geq0$, we solve \eref{cyltrapping} for $R_c< \rho< r$:
\begin{equation}\label{eq:xqpinvortex}
	\frac{x_\qp}{x^0_\qp} \leq 1+\frac{P}{2\pi D}\left[\frac{1}{2}+\ln\left(\frac{\rho}{R_c}\right)\right] < 1.01
\end{equation}
where we assume $r\sim 1$ $\mu$m to 1 mm as the practical (typical) length scale for an area of a film trapping a single vortex ($r\approx80$ $\mu$m for the pads in Type B devices) and $R_c\sim\xi\sim$ 100 nm.  Therefore the QP density dip at the position of the vortex is no more than $\sim P/D$ (or 1\% at most) in relative depth.   Due to the weak logarithmic dependence on length scales, this conclusion is very insensitive to the microscopic model of vortices. 

In the case of multiple vortices in the pad, as long as magnetic field and the quality of the film are roughly homogeneous, vortices are evenly distributed in the pad.  At all magnetic fields one can divide the pad into sub-areas each containing one vortex and has aspect ratio of order of unity, and the spatial distribution of quasiparticle remains approximately uniform for the entire pad.

This approximate homogeneity of the QP density in extended 2D geometries is a direct consequence of the much shorter diffusion time scale $L^2/D$ than the trapping time scale $A/(NP)$ (where $L$ is the typical length between an arbitrary point of the superconducting film and its nearest vortex, and $A$ is the total area of the film).  In extended 2D geometries, $L^2\sim A/N$.   However, in quasi-1D geometries such as the thin connection wire between the pads in our Type B devices, it is possible to have $L^2\gg A/N$, resulting in $L^2/D\gtrsim A/(NP)$.  Therefore large gradient of quasiparticle density can be present along the wire (when the total trapping power in the pads is sufficiently large), while the distribution of quasiparticles in the pads remains approximately homogeneous.

\section{Transmon frequency shift due to quasiparticles}

\begin{figure}[tbp]
    \centering
    \includegraphics[width=0.36\textwidth]{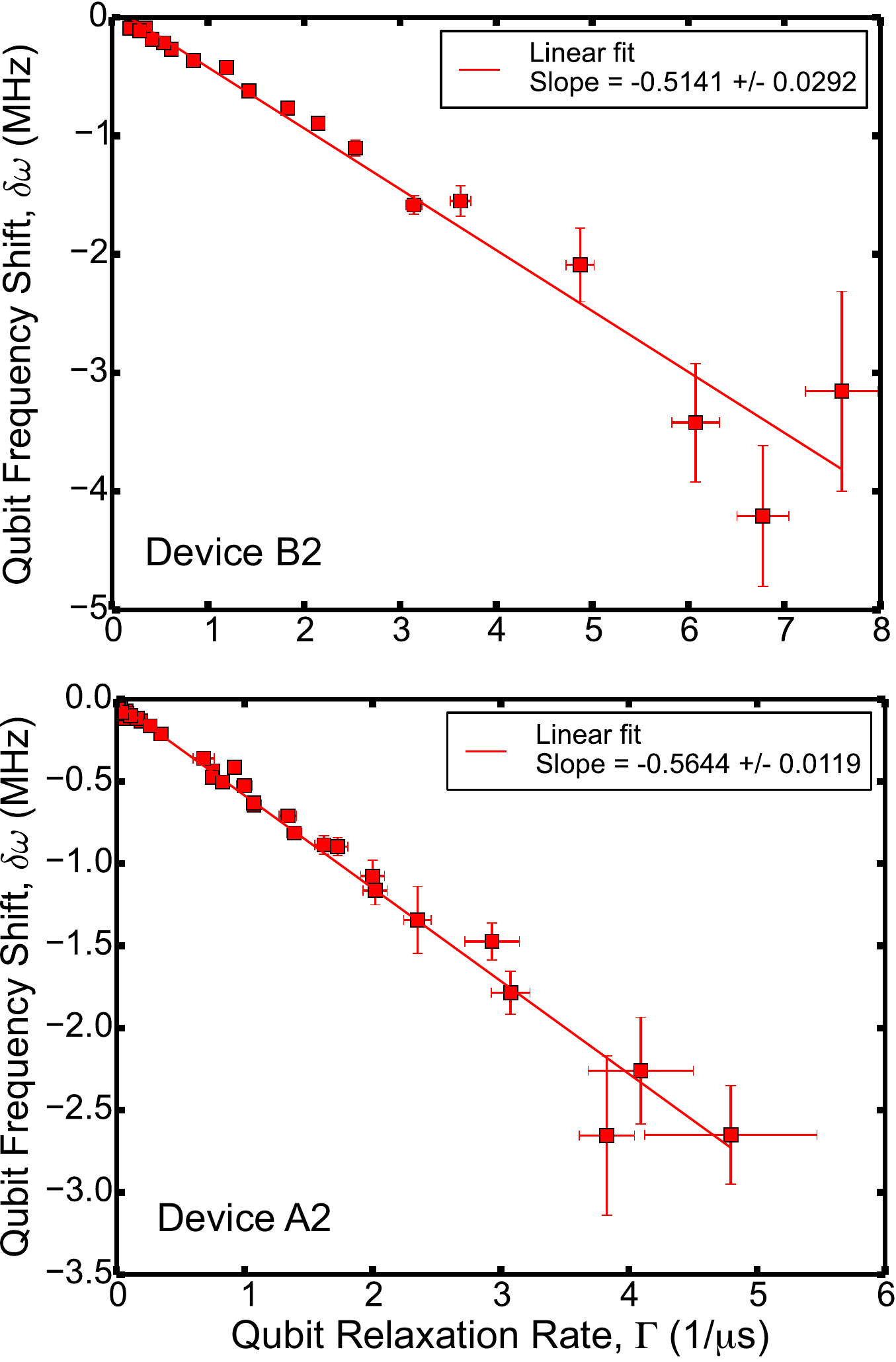}
    \caption{\textbf{Transmon frequency shifts after quasiparticle injection.} Angular frequency shift $\delta \omega$ of two transmon devices after variable delays after quasiparticle injection pulse, plotted as a function of qubit relaxation rate, $\Gamma$ after the same delay. Error bars represent one s.d. statistical uncertainties based on fitting qubit relaxation curves (for $\Gamma$) and Ramsey oscillation (for $\delta\omega$).}
    \label{freqshift}
\end{figure}

In the main text we relied on the proportionality between qubit decay rate and quasiparticle density to study the time evolution of the latter. Indeed, neglecting other relaxation mechanisms, the qubit decay rate $\Gamma$ due to low-energy quasiparticles is
\be\label{qpqd}
\frac{\Gamma}{\omega} = x_\qp \frac{1}{\pi} \sqrt{\frac{2\Delta}{\hbar\omega}}\,,
\ee
where $\omega$ is the qubit frequency and $x_\qp$ the normalized quasiparticle density. Interestingly,
the presence of quasiparticles also results in a change in the qubit frequency, as derived in Refs.~\citen{qp_prl} and~\citen{qp_prb} and measured in a phase qubit in Ref.~\citen{Lenander2011} and a transmon in Ref.~\citen{Paik2011}. Specifically, for a single-junction transmon the frequency change $\delta\omega$ due to low-energy quasiparticles is related to their density by
\be\label{fs}
\frac{\delta \omega}{\omega} = -\frac{1}{2} x_\qp \left[\frac{1}{\pi}\sqrt{\frac{2\Delta}{\hbar\omega}}+1\right]\,.
\ee
The first term in square brackets originates from quasiparticle tunneling events and is a manifestation of fluctu\-ation-dissipation relations: it has the same cause as the one leading to the qubit decay, see \eref{qpqd}. The second term, by contrast, is due to the suppression of the superconducting gap in the presence of quasiparticles

Taken together, \esref{qpqd} and \rref{fs} predict a simple relation between $\delta\omega$ and $\Gamma$:
\be\label{fsd}
\delta \omega = -\frac{1}{2} \Gamma \left[1 + \pi \sqrt{\frac{\hbar\omega}{2\Delta}} \right]\,.
\ee
This relationship can be checked experimentally using our technique: after injecting quasiparticles, we can obtain the transition rate $\Gamma$ by measuring the energy relaxation time $T_1$ of the qubit, as well as the change in its frequency via a $T_2$ Ramsey experiment. As shown in Fig.~\ref{freqshift}, our measurements in two devices display the expected proportionality between $\delta\omega$ and $\Gamma$. However, the slopes of the best-fit lines are lower than those predicted by \eref{fsd} by a factor of approximately 1.7.

\section{Losses due to the vortex flow resistance}

Along with the beneficial effect of trapping nonequilibrium particles, vortices may cause electromagnetic losses in a circuit, if microwaves excite the vortex motion.  However, the current excited by the transmon electromagnetic mode in the vortex-trapping pads is negligible, due to the location of the pads in Type B devices. Therefore, the
vortex-flow loss is minimized.  This is evidenced by the relatively flat qubit $T_1$ vs.~$B$ dependence for a large range of magnetic field (40 mG $<B<$ 200 mG) despite increasing number of vortices.  However, at even higher field ($B\gtrsim$ 200 mG) qubit $T_1$ starts decreasing, which is indicative of losses due to vortices penetrating the gap capacitors where the current density is high.

\begin{figure}[b]
    \centering
    \includegraphics[width=0.40\textwidth]{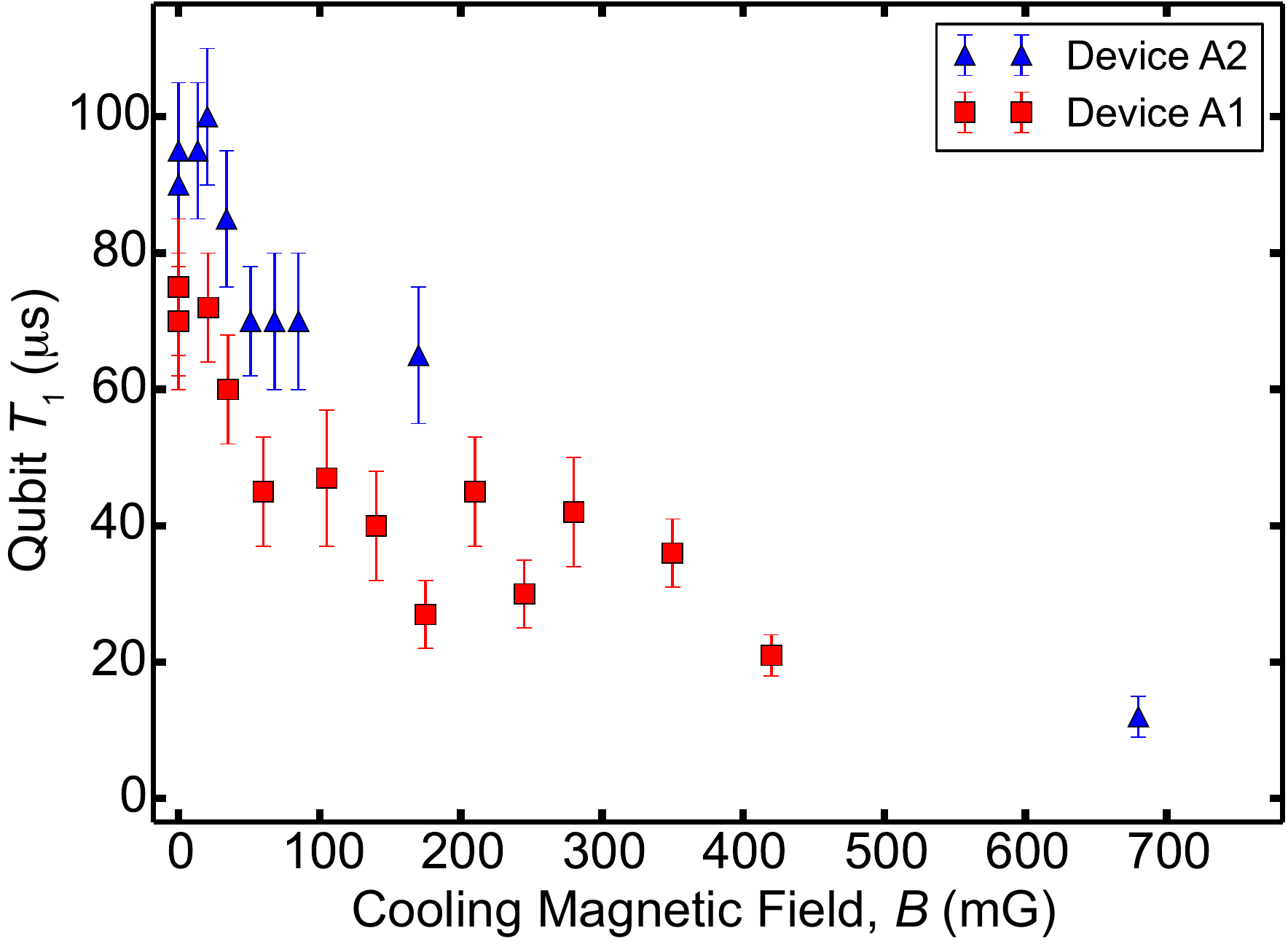}
    \caption{\textbf{Cooling magnetic field dependence of Type A qubit $T_1$.} Qubit relaxation time, $T_1$ as a function of cooling magnetic field for device A1 and A2.  The magnetic field for A2 may be overestimated by up to 50\% in overall scale due to inhomogeneous field for the particular magnetic field coil configuration used for that cool-down.  Error bars represent one s.d. of fluctuations for repetitive measurements (over several hours to one day) within each cycle.}
    \label{condloss}
\end{figure}

We have also studied how the relaxation time, $T_1$, of type A qubits varies with the cooling magnetic field. For two of such qubits (device A1 and A2), we observe no significant changes in $T_1$ at small cooling fields ($\lesssim$ 30mG) and substantial decrease in $T_1$ at large cooling magnetic fields ($\gtrsim$ 30mG) (Fig.~\ref{condloss}).  Because Type A devices have very large electrodes and most likely have trapped vortices when cooled in nominally zero magnetic field, increasing magnetic field (in all regimes of $B$) leads to additional vortices throughout the entire big pads where current density can be either large or small depending on the specific locations.  The decrease in $T_1$ is consistent with additional vortex-flow loss due to increasing number of vortices, and the relatively flat $T_1$ at low magnetic field can be explained as the combined effects of a small enhancement of quasiparticle trapping and a small additional vortex-flow loss.

\begin{table*}[t]
\caption{\textbf{Summary of the parameters of measured transmon devices.}  Repeated cavity entries represent cavities containing two qubits.  Repeated qubit entries represent qubits that have been sequentially mounted into different cavities in different experiment runs (fridge cool-downs).  Quantities unmeasured are left blank.  Note not all devices have been measured under varies cooling magnetic field due to the time-consuming nature of thermal cycling.}
\centering  
\begin{tabular}{c c c c c c c c c c c} 
\hline\hline                        
Qubit	& Cavity	& Cavity Freq	& Qubit Freq	&  $E_C$\footnote{$E_C$ is measured by the transmon anharmonicity (frequency difference between the $|g\rangle\to|e\rangle$ and $|e\rangle\to|f\rangle$ transitions).}	& $E_J$\footnote{$E_J$ is calculated based on the transmon frequency relation $\omega_q=\sqrt{8E_J E_C}-E_C$.}	& $\tss$$_{(B\sim0)}$			& $1/r$$_{(B\sim0)}$	& $T_1$$_{(B\sim0)}$	&  $T_1$$_{(B\sim100mG)}$\\ 
		&		& (GHz)		& (GHz)		& (MHz)	& (GHz)	& (ms)							& (ns)				& ($\mu$s)			& ($\mu$s)\\
\hline
A1		& Al1		& 9.18639		& 6.01059		& 341.0	& 14.79	& 1.5		& 105	& 75\footnote{These are measured during a different cool-down from the data shown in Fig.~2 of the main text (where the average zero-field $T_1$ for the same device is slightly higher).}		& 50$^\textrm{c}$\\
A2		& Al2		& 9.08685		& 6.06365		& 341.8	& 15.01	& 0.25	&NA\footnote{We are unable to quantitatively determine $r$ due to extremely strong trapping.}		& 95		& 65\\
A3		& Al3		& 8.33612		& 6.44094		& 339.0	& 16.95	& 1.6		& 130 	& 75		&\\	
A4		& Al3		& 8.33612		& 6.29598		& 335.0	& 16.41	& 2.8		&		& 70		&\\	
B1		& Al4		& 9.26600		& 5.71200		& 291.7	& 15.45	& 18		& 170	& 9.5		& 25\\
B2		& Al5		& 9.22704		& 5.75975		& 285.8	& 15.99	&		&		& 8.0		& 19\\
B2		& Cu1	& 9.23181		& 5.67777		& 287.1	& 15.49	& 10		& 160	& 7.5		& 16\\
B3		& Al6		& 9.24689		& 5.88310		& 294.6	& 16.19	&		&		& 20		& 30\\
B3		& Cu2	& 9.24613		& 5.89244		& 		&		& 6.5		& 90		& 18		&\\	
B4		& Cu3	& 9.12054		& 6.25856		&		&		& 4.7		& 85		& 16		&\\	
B5		& Cu3	& 9.12054		& 6.61310		&		&		& 13.5	& 80		& 8.5		&\\ [0.1ex] 
\hline 
\end{tabular}
\label{} 
\end{table*}